\documentclass[conference]{IEEEtran}
\usepackage{cite}
\usepackage{amsmath,amssymb,amsfonts}
\usepackage{algorithmic}
\usepackage{graphicx}
\usepackage{textcomp}
\usepackage{xcolor}
\usepackage[hyphens]{url}
\usepackage{todonotes}
\def\BibTeX{{\rm B\kern-.05em{\sc i\kern-.025em b}\kern-.08em
    T\kern-.1667em\lower.7ex\hbox{E}\kern-.125emX}}

\usepackage{cite}
\usepackage{microtype}
\usepackage{subfig} 
\usepackage{tikz}
\usetikzlibrary{positioning}
\usepackage{multirow}
\usepackage{graphicx}
\usepackage{ifthen}
\usepackage{pgfplots}

\usepackage{amssymb}
\usepackage{pifont}

\usepackage{listings}
\usepackage[bookmarks=true,breaklinks=true,letterpaper=true,colorlinks,linkcolor=black,citecolor=blue,urlcolor=black]{hyperref}
\usepackage{breakurl}

\newboolean{showcomments}
\setboolean{showcomments}{true}

\newcommand{\xstate}{{\texttt{xstate}}}
\newcommand{\ustate}{{extra-architectural state}}
\newcommand{\lcm}{LCM}
\newcommand{\lcms}{LCMs}
\newcommand{\toolkit}{\textsc{subrosa}} 
\newcommand{\libsodium}{\texttt{libsodium}}     
\newcommand{\tool}{\textsc{clou}}

\title{Relational Models of Microarchitectures\\for Formal Security Analyses}
\author{
    \IEEEauthorblockN{Nicholas Mosier\IEEEauthorrefmark{1},
    Hanna Lachnitt\IEEEauthorrefmark{1},
    Hamed Nemati\IEEEauthorrefmark{1}\IEEEauthorrefmark{2},
    Caroline Trippel\IEEEauthorrefmark{1}}
    \IEEEauthorblockA{\IEEEauthorrefmark{1}Stanford University\quad\IEEEauthorrefmark{2}CISPA Helmholtz Center for Information Security
    \\\{nmosier, lachnitt, hnemati, trippel\}@stanford.edu}
}

\begin{document}
\maketitle
\thispagestyle{plain}
\pagestyle{plain}

\begin{abstract}
There is a growing need for hardware-software contracts which precisely define the implications of microarchitecture on software security---i.e., \textit{security contracts}.
It is our view that such contracts should explicitly account for \textit{microarchitecture-level implementation details} that underpin hardware leakage, thereby establishing a direct correspondence between a contract and the microarchitecture it represents. At the same time, these contracts should remain as \textit{abstract} as possible so as to support efficient formal analyses.
With these goals in mind, we propose \textit{leakage containment models} (\lcms{})---novel \textit{axiomatic} security contracts which support formally reasoning about the security guarantees of programs when they run on particular microarchitectures.

Our core contribution is an axiomatic vocabulary for formally defining \lcms{}, derived from the established axiomatic vocabulary used to formalize processor \textit{memory consistency models}. Using this vocabulary, we formalize \textit{microarchitectural leakage}---focusing on leakage through hardware memory systems---so that it can be automatically detected in programs.
To illustrate the efficacy of \lcms{}, 
we present two case studies.
First, we demonstrate that our leakage definition faithfully captures a sampling of (transient and non-transient) microarchitectural attacks from the literature.
Second, we develop a 
static analysis tool based on \lcms{} which automatically identifies Spectre vulnerabilities in programs and scales to analyze realistic-sized codebases, like \libsodium{}. 

\end{abstract}

\section{Introduction}
\label{sec:intro}
Hardware which is under-specified or whose implementation deviates from its specification can introduce correctness bugs and/or security vulnerabilities into seemingly correct and secure programs~\cite{spectre:survey,rambleed, amdtlb, ARMHazard, tricheck, guanciale:inspectre, ctfoundations:pldi:20, guanciale:oakland16, pandora:isca:21}.
Unfortunately, \textit{microarchitectural attacks}~\cite{ge:surveycachetimingattacks} expose a notable deficiency in how hardware-software contracts have historically defined \textit{software-visible state}.
Microarchitectural attacks are side/covert channel attacks which enable leakage/communication as a direct result of hardware optimizations.
Thus, rather than consisting solely of state that can be directly accessed with committed user-facing instructions (i.e., architectural state), software-visible state actually also includes any microarchitectural state that can be leaked/communicated via hardware side/covert channels.

Writing secure software in the presence of hardware side/covert channels requires new hardware-software contracts which remedy our inadequate definition of software-visibility. Specifically, \textbf{\textit{security contracts} }should be designed which soundly abstract and expose to software the security implications of particular microarchitectures.
Such contracts can support the design of automated tools that 
\textit{detect} vulnerabilities in programs, \textit{evaluate} hardware and software mitigations, and (optimally) \textit{repair} vulnerable software to render it secure.

\textbf{Hardware-software contracts for security:} One well-established way to counter microarchitectural attacks that manifest as \textit{timing channels} is with \textit{constant-time (CT) programming}---a paradigm that disallows the processing of secrets by \textit{transmit instructions}~\cite{dawg:kiriansky,jiyong:stt} (i.e., transmitters)
which can leak their results, operands, or even data at rest in architectural structures~\cite{pandora:isca:21} via their variable impact on execution time. However, even CT programming requires a type of security contract which precisely identifies transmitters and articulates their leakage implications. Historically, CT programming disallows secret-dependent \textit{branches} and \textit{memory accesses}
~\cite{arm:mbed, libsodium, pornin:whyctc, pornin:cttk}. However, these restrictions are insufficient for modern hardware where secrets can be steered towards \textit{transient} transmitters~\cite{spectre,meltdown,zombieload,ridl,fallout}.
Also, the scope of transmitters extends beyond branch and memory instructions~\cite{pandora:isca:21}.

To address the need for security contracts, various proposals have emerged~\cite{yu:oisa,arm:dit,zagieboylo:iflowisa,woodruff:cheri,formalapproach:csf:19, guanciale:inspectre, heretostay:arxiv:2019, guarnieri:spectector, guarnieri:contracts, neverran:sp:19, ctfoundations:pldi:20, vassena:blade, cat-spectre}.
Some require \textit{hardware enhancements} to explicitly track/enforce contract-level security primitives~\cite{woodruff:cheri,zagieboylo:iflowisa, yu:oisa, arm:dit}.
Other contracts restrict the \textit{scope of hardware features} that they consider when summarizing a microarchitecture's security implications, focusing on in-order~\cite{formalapproach:csf:19, guanciale:inspectre,guarnieri:spectector} and single-core processor designs~\cite{heretostay:arxiv:2019, formalapproach:csf:19, guanciale:inspectre, guarnieri:contracts,neverran:sp:19,guarnieri:spectector,ctfoundations:pldi:20,vassena:blade} for example.
Most recently, security contracts have been proposed which solely expose \textit{transient leakage} through microarchitecture to software~\cite{heretostay:arxiv:2019, neverran:sp:19, formalapproach:csf:19, guarnieri:spectector, ctfoundations:pldi:20, guanciale:inspectre, vassena:blade, cat-spectre} or highly restrict the non-transient leakage they can capture~\cite{guarnieri:contracts}. Notably, some of these contracts require \textit{hard-coding} pre-defined \textit{observations} that a program can produce as it executes on a microarchitecture, given \textit{known side/covert-channels}~\cite{guanciale:inspectre,guarnieri:contracts,cat-spectre,ctfoundations:pldi:20,vassena:blade}.

Existing security contracts share a couple of \textbf{key limitations}. First, their use of \textit{known observations} does not capture the root cause of microarchitectural leakage.
Second, they are not easily verifiable with respect to microarchitecture. The extreme degree to which they abstract away hardware details makes it difficult (if not impossible) to establish a connection between contract primitives and hardware features described with hardware description languages (HDLs). 

\textbf{Our approach:}
Towards resolving the limitations of prior work,  \textbf{our first insight} is that security contracts should explicitly account for \textit{microarchitectural implementation details} that underpin hardware leakage. In doing so, security contracts can be directly related to, and even synthesized from, the microarchitectures they represent~\cite{rtl2uspec:hsiao:2021}.
Moreover, a generic leakage definition can be established which encompasses a wider range of microarchitectural attacks.



Thus, we propose \textbf{\textit{leakage containment models (\lcms{})}}---novel \textit{axiomatic} hardware-software contracts designed to support automatically reasoning about the confidentiality guarantees of programs when they run on particular microarchitectures.
While there has been a particular emphasis in the literature on formalizing security contracts \textit{operationally}, we take an axiomatic approach in this work. In doing so, we expect \lcms{} to benefit from recent work on automatically synthesizing axiomatic specifications of hardware from RTL directly~\cite{rtl2uspec:hsiao:2021}.


\lcms{} are designed to capture the root cause of microarchitectural leakage, described as follows. Specifically, for each
\textit{architecture-level execution} of a program, there may be more than one corresponding \textit{microarchitecture-level execution} that achieves the same software-visible effect.\footnote{In this paper, \textit{software-visibility} is used in the traditional sense to refer to observable program behavior in the absence of hardware side/covert channels.} Furthermore, \textit{which} microarchitecture-level execution is realized when a program runs on a hardware implementation generally depends on the outcome(s) of dynamic microarchitectural information flow(s). If an attacker can distinguish one
microarchitecture-level execution from another, it may infer some function of the data involved in these information flows. As an example, consider a victim program running on a processor with core-private L1 caches. Either an L1 cache \textit{miss} or \textit{hit} (i.e, one of two microarchitecture-level execution possibilities) will occur on behalf of an architecture-level load in the victim program. Moreover, whether a miss or hit occurs depends on the outcome of the load \textit{microarchitecturally reading} the cache state that was \textit{microarchitecturally written} by the last access to the same cache line. A software-based attacker can distinguish these two microarchitecture-level execution scenarios by timing the load's execution latency~\cite{ge:surveycachetimingattacks}, leaking some function of the address bits involved in the culprit microarchitectural information flow.

Given these ingredients for hardware leakage, \textbf{our second insight} is that \lcms{} can \textit{define} and \textit{directly compare} an \textit{architectural semantics} and a \textit{microarchitectural semantics}
for a program
to pinpoint potential hardware-induced program leaks.
A program's architectural semantics encodes the software-visible ways in which it can execute; each execution possibility differs according to the architectural information flows it exhibits. A program's microarchitectural semantics encodes its distinct microarchitectural execution possibilities which differ according to their microarchitectural information flows.
To define \textit{microarchitectural leakage} based on \lcms{}, we first identify which microarchitectural execution of a program is \textit{implied} by each architectural execution possibility in the absence of interference. Then, a program is examined to determine if its microarchitectural semantics can \textit{ever} deviate from what is architecturally-implied. If so, the program is susceptible to hardware leakage. \lcms{} also leverage a program's \textit{speculative semantics} to reason about transient leakage~\cite{guarnieri:contracts}.

In designing \lcms{}, we leverage \textbf{our third insight}---that \textit{memory consistency models} (MCMs)~\cite{lamport:sc,alglave:herd}
define the same sort of architectural program semantics that \lcms{} require.
To summarize, MCMs articulate which architecture-level information flows between shared memory operations in a parallel program are legal; distinct flows constitute distinct architecture-level (i.e., software-visible) program executions. Since established, formally specified MCMs already provide a key building block of \lcms{}---an architectural semantics for programs---we elect to derive \lcms{} from MCMs.
A key benefit of this design choice is that security analyses built on \lcms{} can leverage a rich literature in MCM analysis and verification~\cite{alglave:herd, Alglave:litmus,alglave:dsotf,lustig:automated, Bornholt:memsynth,wickerson:memalloy,alglave:diy,checkweb}.

Overall, this paper lays the foundation for formally evaluating program security from high-level language code down to hardware microarchitectures and makes the following contributions:

\begin{itemize}
    \item \textbf{Axiomatic security contracts:} We propose \lcms{}---novel security contracts---and an axiomatic vocabulary for defining them, derived from axiomatic MCMs. Our formal \lcm{} vocabulary supports advanced processor features like out-of-order and multi-core execution and captures both transient and non-transient leakage.
    
    \item \textbf{Leakage formalization:}
    Using our axiomatic \lcm{} vocabulary, we formalize \textit{microarchitectural leakage}---focusing on leakage through hardware memory systems---so that it can be automatically detected in programs.
    
    \item \textbf{Leakage detection:} 
    First, we demonstrate that our leakage definition faithfully captures a sampling of (transient and non-transient) microarchitectural attacks from the literature~\cite{spectre,jiyong:stt,spectrev4,ctfoundations:pldi:20,pandora:isca:21,spectrev1p1}. 
    Second, we develop a static analysis tool, \tool{}, based on \lcms{} which automatically \textit{identifies} and optimally \textit{repairs} (via fence insertion) \textsc{Spectre v1}~\cite{spectre} and \textsc{Spectre v4}~\cite{spectrev3av4} vulnerabilities in programs. We use \tool{} to analyze 15 \textsc{Spectre v1}~\cite{kocher:sepcterv1-benchmarks} and 14 \textsc{Spectre v4}~\cite{spectrev4-benchmarks} benchmark programs and the \libsodium{} crypto-library~\cite{libsodium}.
    
    \item \textbf{\toolkit{} toolkit:} 
    To support future research, we design the \toolkit{} toolkit, built on top of Alloy~\cite{alloy}, to mechanize \lcms{} and support their formalization. 

    
\end{itemize}

\section{Background and Motivation}
\label{sec:motivating}


\lcms{} define both an \textit{architectural semantics} and a \textit{microarchitectural semantics} for programs.
A program's architectural semantics encodes the distinct software-visible ways in which it can execute; such a semantics is ISA-specific.
A program's microarchitectural semantics encodes the distinct ways in which it can \textit{microarchitecturally execute}; such a semantics is implementation-specific.
Thus, \lcms{} are defined \textit{per-microarchitecture}.
In this section, we discuss how MCMs are specified axiomatically, focusing on the architectural semantics they define for programs which \lcms{} use out-of-the-box.
\subsection{Defining Memory Consistency Models Axiomatically}
\label{subsec:mcmmotiv}
MCMs define the value(s) that can be legally returned by shared-memory loads in a parallel program.
Since MCMs are central to reasoning about parallel program correctness, a large body of work is devoted to formalizing them~\cite{manson:java, boehm:cppconcurrency, petri:cooking, Batty:mathematizingc++, batty:overhauling, nienhuis:c11operational, Wickerson2015, nagarajan2020primer, owens:better, alglave:herd, pulte:armv8, nvidia:ptx, RISCV:rvtso:rvwmo}.
In particular, an axiomatically formalized MCM consists of a \textit{predicate} on \textit{candidate executions} of programs. 

\subsubsection{Event Structures}
\label{sec:event-structures}
A candidate execution of a program is derived from an \textit{event structure}~\cite{alglave:dsotf}, which describes a particular \textit{control-flow path} of the program (with all branches resolved). Precisely, an event structure $E\triangleq(\texttt{MemoryEvent}, \texttt{Location}, \texttt{address}, \texttt{po})$ consists of: 
\begin{itemize}
    \item \texttt{MemoryEvent}: a set of all program instructions that read or write memory.
    \texttt{MemoryEvent} is a subset \texttt{Event}---a set  which contains all program instructions.
    \item \texttt{Location}: a set of all architectural memory locations which are accessed by program \texttt{Read}/\texttt{Write} events---\texttt{Read} and \texttt{Write} are disjoint subsets of \texttt{MemoryEvent}.
    \item \texttt{address}: a binary relation that maps each \texttt{MemoryEvent} to the single \texttt{Location} it accesses.
    \item \texttt{po}: a binary relation that maps each \texttt{Event} to all \textit{committed} \texttt{Events}
    that follow it in \textit{program order}---\texttt{po} is a per-thread total order on committed instructions.
\end{itemize}

\subsubsection{Candidate Executions}
\label{sec:candidate-executions}
An event structure for a program can be extended to a set of candidate executions, each of which differs with respect to the shared memory interactions between instructions that are realized.
Concretely, we can complete an event structure to produce a candidate execution by adding an \textit{execution witness} $X\triangleq(\texttt{rf}, \texttt{co}, \texttt{fr})$, which is comprised of three new relations involving \textit{same-address} \texttt{MemoryEvents}:
\begin{itemize}
    \item \texttt{rf} (\textit{reads-from}): a binary relation that maps each \texttt{Write} to all same-address \texttt{Reads} that read from it.
    \item \texttt{co} (\textit{coherence-order)}: a binary relation that maps each \texttt{Write} to all same-address \texttt{Writes} that follow it in coherence order.
    \item \texttt{fr} (\textit{from-reads}): a binary relation that maps each \texttt{Read} to all \texttt{co}-successors of the \texttt{Write} that it read from.  \texttt{fr$=\sim$rf.co}, where $\sim$ is relational transpose and . is relational join.
\end{itemize}
Collectively, \texttt{rf}, \texttt{co}, and \texttt{fr} comprise the \texttt{com} (\textit{communication}) relation---$\texttt{com} = \texttt{rf} + \texttt{co} + \texttt{fr}$, where $+$ is set union.

\subsubsection{Consistency Predicates}
\label{sec:consistency-predicates}
A candidate execution is uniquely defined by an event structure $E$ and an execution witness $X$. An MCM is then defined by a \textit{consistency predicate} which renders candidate executions consistent (allowed) or inconsistent (disallowed) with respect to it. In constructing this predicate, axiomatic MCM specifications often consider a wider range of events (e.g., fences) and relations, such as:
\begin{itemize}
    \item \texttt{ppo}: a binary relation that maps an \texttt{Event} to a \texttt{po}-later \texttt{Event} if the ISA guarantees they will be executed in order \textit{from the perspective of all cores} in the shared memory system.
    \item \texttt{fence}: a binary relation that maps an \texttt{Event} $e_0$ to another \texttt{Event} $e_1$ if $e_0$ is ordered before $e_1$ by an explicit \textit{synchronization event} (e.g., a fence/barrier).
\end{itemize}

For MCMs which do not order \texttt{Reads} with \texttt{po}-later \texttt{MemoryEvents} by default (i.e., via \texttt{ppo}), a \texttt{dep} (\textit{dependency}) relation is used to selectively enforce these orders. \texttt{dep} encodes syntactic dependencies \textit{through registers}, and is comprised of the following three sub-relations:
\begin{itemize}
    \item \texttt{addr} (\textit{address dependency}): a binary relation that maps a \texttt{Read} to \texttt{po}-subsequent \texttt{MemoryEvent} when the \texttt{Location} accessed by the \texttt{MemoryEvent} depends syntactically on the value returned by the \texttt{Read}.
    \item \texttt{data} (\textit{data dependency}): a binary relation that maps a \texttt{Read} to a \texttt{po}-subsequent \texttt{Write} when the written value depends syntactically on the value read.
    \item \texttt{ctrl} (\textit{control dependency}): a binary relation that maps a \texttt{Read} to a \texttt{po}-subsequent \texttt{MemoryEvent} when the control flow decision of whether to execute the \texttt{MemoryEvent} depends syntactically on the value read.
\end{itemize}

An example consistency predicate defines the Total Store Order (TSO) MCM used by Intel x86 processors~\cite{intel:x86}. It is composed of the conjunction of three auxiliary predicates---\textit{sc\_per\_loc}, \textit{rmw\_atomicity}, and \textit{causality}~\cite{alglave:herd}. Below we define the most relevant two for the ideas presented in this paper:
\begin{itemize}
    \item \textit{{sc\_per\_loc}}: \textit{{\tt\{rf~+~co~+~fr~+~po\_loc\}} is acyclic}, where \texttt{po\_loc} is the subset of \texttt{po} that relates same-address \texttt{MemoryEvents}.
    \item \textit{causality}:\textit{ \texttt{\{{rfe} + {co} + {fr} + {ppo} + {fence}\}} is acyclic}. 
     For x86-TSO, \texttt{ppo} includes all $\texttt{Write}\rightarrow\texttt{Write}$ and $\texttt{Read}\rightarrow\texttt{MemoryEvent}$ tuples in $\texttt{po}$.
    \texttt{rfe} (\textit{reads-from external}) the subset of \texttt{rf} that relates \texttt{Events} on different threads.
\end{itemize}

\subsection{An Axiomatic Architectural Semantics for \lcms{}}
Recall that
\lcms{} define an ISA-specific \textit{architectural semantics}, which encodes the various software-visible ways in which programs can execute; each execution possibility differs according to the architectural information flows it exhibits.
Now consider an ISA MCM, defined axiomatically with the help of a consistency predicate.
Notably, the \texttt{com} relation (\S\ref{sec:candidate-executions}) encodes architectural information flows through shared memory for a specific candidate execution.
Thus, for a given program, its set of \textit{consistent candidate executions}---i.e., those candidate executions which are consistent with the consistency predicate---constitute its architectural semantics as required by \lcms{}.

More precisely, consistent candidate executions comprise a program's architectural semantics \textit{restricted to memory instructions}.
In this paper, we use \lcms{} to model leakage on behalf of hardware memory systems optimizations, particularly cache optimizations. Hence, this restriction is appropriate.

\section{Leakage Containment Models}
\label{sec:makingacaselcm}
\subsection{What Memory Models are Missing}
\label{sec:mcmsfail}
Recent work identifies similarities between MCMs and the sorts of security contracts that software and hardware designers would benefit from~\cite{checkmate, cat-spectre, neverran:sp:19}.
However, MCMs themselves
do not offer a complete security contract solution.
To demonstrate why, consider the classic \textsc{Spectre v1}~\cite{spectre} program in Fig.~\ref{fig:spectre-v1-C} and its corresponding assembly pseudo-code in Fig.~\ref{fig:spectre-v1-asm}. Due to the branch, axiomatic MCM definitions would consider two distinct event structures for this program---one which corresponds to the \textit{not-taken} branch outcome (Fig.~\ref{fig:not-taken-event-structure}) and the other which corresponds to the \textit{taken} outcome (Fig.~\ref{fig:taken-event-structure}).
Note that axiomatic MCM definitions facilitate modeling both event structures and candidate executions as \textit{directed graphs} where nodes are \texttt{MemoryEvents} labeled with the \texttt{Location} they access---per the \texttt{address} relation---and edges denote types of ``happens-before''~\cite{lamport:happensbefore} (i.e., sequencing) relationships---per relations such as \texttt{po}, \texttt{com}, and \texttt{dep}.

Each event structure in Fig.~\ref{fig:motivating} can be extended to \textit{exactly one} candidate execution. Thus, there are two possible candidate executions for \textsc{Spectre v1}. This is because every memory access in the \textsc{Spectre v1} program touches a \textit{distinct} memory location; thus, only one instantiation of the \texttt{com} relation is possible for each event structure. Specifically, all \texttt{Read} events read from the \textit{initial state} of memory---also by convention, no \texttt{rf} edges are explicitly drawn since initialization writes are not explicitly modeled. Second, the sole \texttt{Write} event is coherence-ordered after the last \textit{initialization write} to the same memory location---by convention, no \texttt{co} edges are drawn. Without \texttt{rf} and \texttt{co} edges, there are no \texttt{fr} edges (\S\ref{sec:candidate-executions}). Figs.~\ref{fig:not-taken-event-structure}~and~\ref{fig:taken-event-structure} thus \textit{also} constitute candidate executions. Moreover, they constitute \textit{consistent candidate executions} according to TSO (\S\ref{sec:consistency-predicates}) making them valid architectural execution possibilities on Intel processors.
Fig.~\ref{fig:taken-event-structure} uses gray edges to depict instances of the \texttt{dep} relation, although it is not a distinguishing feature of event structures or candidate executions. 

As is known, the program in Fig.~\ref{fig:spectre-v1-C} exhibits a variety of hardware-induced leakage when run on modern processors.
First, the \textit{addresses} accessed by instructions 1 and 2 in Fig.~\ref{fig:not-taken-event-structure} and instructions 1, 2, 5, 6, and 7 in Fig.~\ref{fig:taken-event-structure} may be leaked to an attacker via a simple cache side-channel attack.
Second, the \textit{data} returned by read instructions 2 and 5 in Fig.~\ref{fig:taken-event-structure} can be leaked. This is because the \texttt{addr} dependency from instruction 2 (resp. 5) to 5 (resp. 6) indicates that the data returned by 2 (resp. 5) is supplied as the address operand of 5 (resp. 6), an instruction which we established can leak its address operand.
Third, the outcome of the branch can be leaked.
Moreover, the program in Fig.~\ref{fig:spectre-v1-C} exhibits speculative leakage which cannot be discerned from Figs.~\ref{fig:not-taken-event-structure}~and~\ref{fig:taken-event-structure}.
In summary, MCMs cannot directly capture microarchitectural leakage out-of-the-box.




\newsavebox\spectrevone
\begin{lrbox}{\spectrevone}
\begin{minipage}[b]{.4\linewidth}
\begin{lstlisting}
if(y < size_A)
  x = A[y];
  tmp &= B[x];
\end{lstlisting}
\end{minipage} 
\end{lrbox}

\newsavebox\spectrevoneasm
\begin{lrbox}{\spectrevoneasm}
\begin{minipage}[b]{.4\linewidth}
\begin{lstlisting}
R size ->r1
R y ->r2 
r3 <-(r2<r1)
BEQZ r3, 8
R A+r2 ->r4
R B+r4 ->r5
W tmp <-tmp&r5
skip
\end{lstlisting}
\end{minipage} 
\end{lrbox}

\begin{figure}[t]
\begin{minipage}[b]{.44\linewidth}
\vspace{2pt}
\subfloat[\textsc{Spectre v1}]{
        \label{fig:spectre-v1-C}
        \usebox\spectrevone
        }
        
\subfloat[\textsc{Spectre v1} assembly pseudo-code]{
    \label{fig:spectre-v1-asm}
        \usebox\spectrevoneasm
}
\end{minipage}
\begin{minipage}[b]{.48\linewidth}
\subfloat[\textit{Not-Taken} event structure \textit{and} candidate execution]{
    \label{fig:not-taken-event-structure}
    \begin{minipage}[b]{\linewidth}
    \centering
    \footnotesize
    \begin{tikzpicture}[->,>=stealth,shorten >=1pt,auto,node distance=.62cm, node/.style={rectangle,draw=none,fill=none,minimum size=1mm}]
        \footnotesize
        \node[node] (i0) {1: \textbf{R} size $\rightarrow$ r1};
        \node[node] (i1) [below of=i0] {2: \textbf{R} y $\rightarrow$ r2};

        \path[every node/.style={font=\sffamily\footnotesize, fill=none,inner sep=2pt}]
                (i0) edge [left] node {po} (i1)
                ;
                
        \end{tikzpicture}
    \end{minipage} 
}

\subfloat[\textit{Taken} event structure \textit{and} candidate execution]{
    \label{fig:taken-event-structure}
    \begin{minipage}[b]{\linewidth}
    \centering
    \footnotesize
    \begin{tikzpicture}[->,>=stealth,shorten >=1pt,auto,node distance=.62cm, node/.style={rectangle,draw=none,fill=none,minimum size=1mm}]
        \footnotesize
        \node[node] (i0) {1: \textbf{R} size $\rightarrow$ r1};
        \node[node] (i1) [below of=i0] {2: \textbf{R} y $\rightarrow$ r2};
        \node[node] (i4) [below of=i1] {5: \textbf{R} A+r2 $\rightarrow$ r4};
        \node[node] (i5) [below of=i4] {6: \textbf{R} B+r4 $\rightarrow$ r5};
        \node[node] (i6) [below of=i5] {7: \textbf{W} tmp $\leftarrow$ tmp \& r5};
        
        \coordinate[below right of=i1, yshift=.35cm, xshift=.1cm] (i1-bottom-right);
        \coordinate[below right of=i4, yshift=.35cm, xshift=.2cm] (i4-bottom-right);
        \coordinate[below right of=i5, yshift=.35cm, xshift=.2cm] (i5-bottom-right);
    
        \path[every node/.style={font=\sffamily\scriptsize, fill=none,inner sep=2pt}]
                (i0) edge [left] node {po} (i1)
                (i1) edge [left] node {po} (i4)
                (i4) edge [left] node {po} (i5)
                (i5) edge [left] node {po} (i6)
                (i1.west) edge [gray, bend right=90, left] node {ctrl} (i4.west)
                (i1.east) edge [gray, bend left=90, right, near end] node {ctrl} (i5.east)
                (i1.east) edge [gray, bend left=90, right, near end] node {ctrl} (i6.east)
                (i0.east) edge [gray, bend left=90, right] node {ctrl} (i4.east)
                (i0.west) edge [gray, bend right=90, left, near end] node {ctrl} (i5.west)
                (i0.west) edge [gray, bend right=90, left, near end] node {ctrl} (i6.west)
                (i1-bottom-right) edge [gray, right] node {addr} (i4)
                (i4-bottom-right) edge [gray, right] node {addr} (i5)
                (i5-bottom-right) edge [gray, right] node {data} (i6)
                ;
                
        \end{tikzpicture}
    \end{minipage} 
}
\end{minipage}
\caption{\textsc{Spectre v1} in \protect\subref{fig:spectre-v1-C}, produces two event structures (\S\ref{sec:event-structures})---\protect\subref{fig:not-taken-event-structure} and \protect\subref{fig:taken-event-structure}. Each event structure can be completed with a single execution witness (\S\ref{sec:candidate-executions}). The resulting two candidate executions look identical to the event structures, since no explicit \texttt{com} edges are instantiated (\S\ref{sec:makingacaselcm}).}
\label{fig:motivating}
\end{figure}


\subsection{An Axiomatic Microarchitectural Semantics}
\label{subsec:whylcms}


\lcms{} facilitate reasoning about hardware-induced leakage in programs by augmenting the architectural semantics provided by axiomatic MCMs with a \textit{microarchitectural semantics} that describes the various ways in which a program can microarchitecturally execute. Each execution possibility differs according to microarchitectural information flows it exhibits.

In defining a microarchitectural semantics for \lcms{} we leverage \textit{two key building blocks}, featured in Fig.~\ref{fig:spectrev1-uarch-sem}. Fig.~\ref{fig:spectrev1-uarch-sem} effectively merges together the two \textsc{Spectre v1} candidate executions  (Figs.~\ref{fig:not-taken-event-structure}~and~\ref{fig:taken-event-structure}) into a single graph and adds some new nodes and edges. Some instructions are also omitted for clarity, but numeric instruction labels are retained.

In Fig.~\ref{fig:spectrev1-uarch-sem}, $\top{}$ represents \textit{explicitly} the set of architectural/microarchitectural writes that initialize relevant architectural/microarchitectural state. $\bot$ represents a set of \textbf{\textit{observer}} accesses that observe aspects of final architectural/microarchitectural state after the program runs to completion. In this paper, we assume that the observer ($\bot$) does not share memory with the executing program, and thus it \textit{cannot} interact with the program architecturally (i.e. via \texttt{com}). $\bot$ may only be involved in a \texttt{com} relation with $\top$. However, $\bot$ \textit{can} interact with the program microarchitecturally, such as by probing cache sate. Thus, $\top$ can be involved in \texttt{comx} (\S\ref{sec:comx}) relations with program instructions. Each straight-line path through \texttt{po} edges from $\top$ to $\bot$, together with the \texttt{com} relation, denotes a distinct candidate execution.
In Fig.~\ref{fig:spectrev1-uarch-sem}, the \texttt{com} relation has been explicitly drawn---i.e., note the presence of \texttt{rf} edges in contrast to Figs.~\ref{fig:not-taken-event-structure}~and~\ref{fig:taken-event-structure} which do not model initialization writes. Edges missing a source node are implicitly related to $\top$.


\subsubsection{Modeling Microarchitectural State}

\label{sec:xstate}
The microarchitectural semantics defined by \lcms{} explicitly considers \textit{microarchitectural state}, effectively denoting \textit{which} state elements in a processor are accessed on behalf of architectural program instructions and \textit{how} they are accessed. We refer to said state as  \textit{\ustate{}} (or \xstate{}), meaning that it can consist of any \textit{non-architectural} state in a microarchitecture.\footnote{The term \textit{extra-architectural state} was coined in prior work~\cite{lowepower:xstate}; however, we assign it a different meaning in this paper.}
Fig.~\ref{fig:spectrev1-uarch-sem} illustrates that \xstate{} elements \texttt{s$_0$}, \texttt{s$_1$}, and \texttt{s$_2$} are accessed on behalf of \texttt{Read} instructions 2, 5, and 6, respectively. 
Furthermore, all three \xstate{} accesses are \textit{microarchitectural} read-modify-write operations, denoted by ``RW'' before the \xstate{} identifier in the figure. In other words, \textit{R~y~(RW~$s_0$)~$\rightarrow$~r2} means that architectural \texttt{Read} event $R$, which accesses architectural \texttt{Location} $y$, \textit{induces} a microarchitectural read-modify-write of \xstate{} element $s_0$.

The \xstate{} identifiers used by \lcms{}, such as those featured in Fig.~\ref{fig:spectrev1-uarch-sem},
may represent a \textit{set} of hardware state elements in a microarchitecture.
Furthermore, an instruction can access a \textit{vector} of \xstate{} rather than a single \xstate{} element. Crucially, instructions which access common \xstate{} elements are capable of \textit{communicating} microarchitecturally---modeled by a new communication relation, \texttt{comx} (\S\ref{sec:comx}). In fact, the sole reason for modeling \xstate{} is to establish \texttt{comx} for a given candidate execution.
Instructions may also access different \xstate{} elements in different ways depending on execution context, as described below.

In this paper, we seek to model hardware leakage due to memory systems optimizations, particularly cache optimizations. Thus, we consider \xstate{} accessed on behalf of architectural \textit{memory instructions} only.
In particular, the \xstate{} elements we model in this paper are intended to capture the ways in which \textit{same-core memory instructions} can communicate \textit{microarchitecturally}---said \xstate{} then effectively represents the core-private cache lines and store buffer entries that are accessed on behalf of architectural memory instructions.

To understand what the above \xstate{} modeling choice means for \textit{how} memory instructions access these abstract \xstate{} elements, consider the following.
In general, (cacheable) architectural read instructions either microarchitecturally read a cache line (a cache hit) or microarchitecturally read-modify-write a cache line (a cache miss). With respect to a local store buffer, architectural reads \textit{may} microarchitecturally read (i.e., forward) data from a pending store. Similarly, (cacheable) architectural writes always behave as cache line read-modify-writes, unless they are executing on a microachitecture with a no-write-allocate cache policy.
With respect to store buffer state, stores always behave as microarchitectural writes. 
Given \xstate{} elements which collectively represent the core-private cache line and store buffer entries accessed on behalf of an architectural memory instruction: \textit{read hits} read \xstate{} (from the cache or from a pending store in the store buffer), \textit{read misses} read-modify-write \xstate{} (namely a cache line), and \textit{writes} read-modify-write \xstate{} (namely a cache line which subsumes the store buffer write).



\subsubsection{Modeling Microarchitectural Information Flow}
\label{sec:comx}
Fig.~\ref{fig:spectrev1-uarch-sem} shows that \lcms{} define a \texttt{comx} relation which lifts \texttt{com}~\cite{alglave:herd} to \xstate{} accesses; \texttt{com} relates same-\texttt{address} operations, while \texttt{comx} relates same-\xstate{} operations.
Recall that \lcms{} use the \texttt{com} relation to encode the architectural information flows that distinguish program executions according to their architectural semantics. Likewise, \lcms{} use the \texttt{comx} relation to encode microarchitectural information flows that distinguish program executions according to their microarchitectural semantics. Just as a consistency predicate was used to rule out illegal instantiations of \texttt{com}, a similar \textit{confidentiality predicate} must be defined to rule out illegal instantiations of \texttt{comx} according to a specific hardware implementation. \S\ref{sec:specexamples} discusses features of confidentiality predicates that are required to capture different sorts of known hardware optimizations.  




\subsubsection{Modeling Microarchitectural Leakage}
\label{sec:modeling-leakage}
\lcms{} formalize \textit{microarchitectural leakage} by (1) determining which microarchitectural semantics (\texttt{comx} edges) are implied by a given architectural semantics (\texttt{com} edges), and (2) detecting when a program's microarchitectural semantics deviates from architectural expectation.

For example, consider an \texttt{rf} edge which relates a write to a \textit{same-core} read that it sources---called \texttt{rf}-internal or \texttt{rfi}~\cite{alglave:herd}. Consider also our \xstate{} of interest which corresponds to core-private processor cache lines or store buffer entries.
In the absence of interference, an \texttt{rfi} edge which relates some \texttt{Write} $w$ to some \texttt{Read} $r$, implies a consistent \texttt{rfx} edge---an \texttt{rfx} edge which relates $w$ to $r$. 
In other words, if non-interference holds, a read $r$ which \textit{architecturally} reads from a same-core write  $w$ will further \textit{microarchitecturally} read from the core-private cache line or buffer entry populated by $w$.
If $r$ reads from a cache line populated by a different instruction, this means that the cache line was evicted by an interfering access in between $r$'s and $w$'s cache accesses. If $r$ forwards data from an interfering store residing in the store buffer rather than reading from $w$, then it exhibits memory address mis-speculation (\S\ref{subsec:whyspeculation}) which will be eventually rolled back.

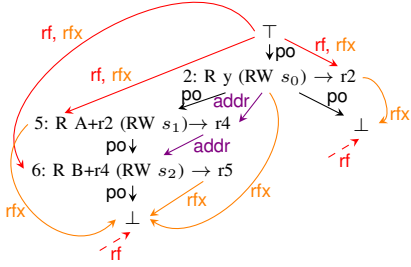
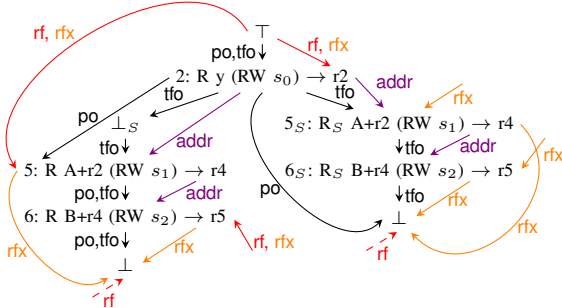
\begin{figure}[t!]
 \subfloat[The \textit{microarchitectural semantics} of \lcms{} captures communication between instructions via \xstate{} ($s_0$, $s_1$, and $s_2$).]{
    \label{fig:spectrev1-uarch-sem}
        \begin{minipage}[b]{.96\linewidth}
        \vspace{-20pt}
        \centering
        \begin{tikzpicture}[->,>=stealth,shorten >=1pt,auto,node distance=.62cm, node/.style={rectangle,draw=none,fill=none,minimum size=1mm}]
            \scriptsize
            \node[node] (top) {$\top$};
            \node[node] (Ry) [below of=top] {2: R y (RW $s_0$) $\rightarrow$ r2};
            \node[node] (space-left) [left of=Ry, xshift=-1.2cm] {};
            \node[node] (space-right) [right of=Ry, xshift=.6cm] {};
            \node[node] (RA) [below of=space-left] {5: R A+r2 (RW $s_1$)$\rightarrow$ r4};
            \node[node] (RB) [below of=RA] {6: R B+r4 (RW $s_2$) $\rightarrow$ r5};
            \node[node] (bottom-left) [below of=RB] {$\bot$};
            \node[node] (bottom-right) [below of=space-right] {$\bot$};
            
            \coordinate[above right of=Ry, xshift=.5cm] (Ry-xinit);
            \coordinate[above left of=RA, xshift=-1.3cm] (RA-init);
            \coordinate[below left of=RB, xshift=-1.2cm] (RB-init);
            \coordinate[above of=RA, xshift=.5cm] (RA-xinit);
            \coordinate[above right of=RB, xshift=.5cm] (RB-xinit);
            \coordinate[below left of=bottom-right] (bottom-right-obs);
            \coordinate[below left of=bottom-left] (bottom-left-obs);
            \coordinate[below left of=Ry] (Ry-init);
            \coordinate[below right of=RB, yshift=.3cm, xshift=.5cm] (RB-bottom-right);
            \coordinate[below right of=RA, yshift=.3cm, xshift=.5cm] (RA-bottom-right);
            \coordinate[above right of=Ry, yshift=-.3cm, xshift=.5cm] (Ry-top-right);
            \coordinate[above left of=RA, yshift=-.3cm, xshift=-.5cm] (RA-top-left);
            
            \path[every node/.style={font=\sffamily\scriptsize, fill=none,inner sep=2pt}]
                (top) edge [right] node {po} (Ry)
                (RA) edge [left] node {po} (RB)
                (RB) edge [left] node {po} (bottom-left)
                (Ry) edge [left, near start] node {po~~} (RA)
                (Ry) edge [right, near start] node {~~po} (bottom-right)
                (Ry.south) edge [violet, left, xshift=-.1cm, near start] node {addr} (RA.east)
                (RA-bottom-right) edge [violet, right] node {~addr} (RB)
                (top) edge [red,  right, near start] node {~~rf, {\color{orange}rfx}} (Ry-top-right)
                (top) edge [red, left] node {rf, {\color{orange}rfx}~~~~~} (RA-top-left)
                (top.west) edge [red, bend right=90, left, looseness=1] node {rf, {\color{orange}rfx}~~~} (RB.west)
                (RB-bottom-right) edge [orange, right] node {~rfx} (bottom-left.east)

                (RA.west) edge [orange, bend right=90, looseness=1.5, left] node {rfx} (bottom-left.west)
                (Ry.east) edge [orange, bend left=80, right, near end] node {rfx} (bottom-right.east)
                (Ry.south) edge [orange, bend left=80, right] node {rfx} (bottom-left.east)
                (bottom-left-obs) edge [red, below, dashed] node {rf} (bottom-left.south)
                (bottom-right-obs) edge [red, below, dashed] node {rf} (bottom-right.south)
                ;
        \end{tikzpicture}\vspace{-5pt}
        \end{minipage}  
        }\vspace{-30pt}
        
\subfloat[The \textit{speculative semantics} of \lcms{} demonstrates that leakage can involve speculatively-executed source instructions, denoted with subscript $S$.]{
    \label{fig:spectrev1-spec-sem}
        \begin{minipage}[b]{.96\linewidth}
        \centering
        \begin{tikzpicture}[->,>=stealth,shorten >=1pt,auto,node distance=.62cm, node/.style={rectangle,draw=none,fill=none,minimum size=1mm}]
            \scriptsize
            \node[node] (top) {$\top$};
            \node[node] (Ry) [below of=top] {2: R y (RW $s_0$) $\rightarrow$ r2};
            \node[node] (space-left) [left of=Ry, xshift=-1.2cm] {};
            \node[node] (space-right) [right of=Ry, xshift=1.2cm] {};
            \node[node] (bottom-spec) [below of=space-left] {$\bot{_S}$};
            \node[node] (RA) [below of=bottom-spec] {5: R A+r2 (RW $s_1$) $\rightarrow$ r4};
            \node[node] (RB) [below of=RA] {6: R B+r4 (RW $s_2$) $\rightarrow$ r5};
            \node[node] (bottom-left) [below of=RB] {$\bot$};
            \node[node] (RA-spec) [below of=space-right] {5$_S$: R$_S$ A+r2 (RW $s_1$) $\rightarrow$ r4};
            \node[node] (RB-spec) [below of=RA-spec] {6$_S$: R$_S$  B+r4 (RW $s_2$) $\rightarrow$ r5};
            \node[node] (bottom-right) [below of=RB-spec] {$\bot$};
            
            \coordinate[above right of=Ry, xshift=.5cm] (Ry-xinit);
            \coordinate[above left of=RA, xshift=-.5cm] (RA-init);
            \coordinate[below left of=RB, xshift=-1.2cm] (RB-init);
            \coordinate[above right of=RA-spec, xshift=.5cm] (RA-spec-init);
            \coordinate[above right of=RB-spec, xshift=.5cm] (RB-spec-init);
            \coordinate[below right of=RA, xshift=1.25cm] (RA-xinit);
            \coordinate[below right of=RB, xshift=1.25cm] (RB-xinit);
            \coordinate[below left of=bottom-right] (bottom-right-obs);
            \coordinate[below left of=bottom-left] (bottom-left-obs);
            \coordinate[below left of=Ry] (Ry-init);
            \coordinate[below right of=RB, yshift=.3cm, xshift=.5cm] (RB-bottom-right);
            \coordinate[below right of=RB-spec, yshift=.3cm, xshift=.5cm] (RB-spec-bottom-right);
            \coordinate[below right of=RA, yshift=.3cm, xshift=.5cm] (RA-bottom-right);
            \coordinate[below right of=RA-spec, yshift=.3cm, xshift=.5cm] (RA-spec-bottom-right);
            \coordinate[above left of=RA, yshift=-.3cm, xshift=-.7cm] (RA-top-left);
            \coordinate[above right of=RA-spec, yshift=.1cm, xshift=.4cm] (RA-spec-top-right);
            \coordinate[above right of=RB-spec, xshift=1.5cm] (RB-spec-top-right);
            
            \path[every node/.style={font=\sffamily\scriptsize, fill=none,inner sep=2pt}]
                (top) edge [left] node {po,tfo} (Ry)
                (Ry) edge [left] node[xshift=-4pt, yshift=2pt, near start] {tfo} (bottom-spec)
                (Ry) edge [right] node[xshift=4pt, yshift=2pt, near start] {tfo} (RA-spec)
                (bottom-spec) edge [left] node {tfo} (RA)
                (RA) edge [left] node {po,tfo} (RB)
                (RB) edge [left] node {po,tfo} (bottom-left)
                (Ry.west) edge [left] node {po} (RA-top-left)
                (RA-spec) edge [right] node {tfo} (RB-spec)
                (RB-spec) edge [right] node {tfo} (bottom-right)
                (Ry.south) edge [bend right=90, left] node {po~} (bottom-right.west)
                (top) edge [red,  right, near start] node {~~rf, {\color{orange}rfx}} (Ry-top-right)
                (Ry) edge [violet, right, near end] node {addr} (RA)
                (Ry.east) edge [violet] node {addr} (RA-spec)
                (RA-bottom-right) edge [violet, right] node {~addr} (RB)
                (RA-spec-bottom-right) edge [violet, right] node {~addr} (RB-spec)
                (RB-xinit) edge [red, right, near start] node {rf,  {\color{orange}rfx}} (RB.east)
                (RA-spec-top-right) edge [orange, right] node {~rfx} (RA-spec)
                (RB-spec-top-right) edge [orange, right] node {~rfx} (RB-spec.east)
                (top.west) edge [red, bend right=90, left] node {rf, {\color{orange}rfx}~~~} (RA.west)
                (RB-spec-bottom-right) edge [orange, right] node {rfx} (bottom-right.east)
                (RA-spec.east) edge [orange, bend left=90, right, looseness=1.5] node {rfx} (bottom-right)
                (RB-bottom-right) edge [orange, right] node {rfx} (bottom-left.east)
                (RA.west) edge [orange, bend right=80, left] node {rfx} (bottom-left)
                (bottom-right-obs) edge [red, below, dashed] node {rf} (bottom-right.south)
                (bottom-left-obs) edge [red, below, dashed] node {rf} (bottom-left.south)
                ;
        \end{tikzpicture}\vspace{-5pt}
        \end{minipage}  
    }    
\caption{\lcms{} extend MCMs with a \textit{microarchitectural semantics}, as in \protect\subref{fig:spectrev1-uarch-sem}---to modeling microarchitectural leakage---and a \textit{speculative semantics}, as in \protect\subref{fig:spectrev1-spec-sem}, to model transient leakage.
}
\label{fig:spectrev1}
\end{figure}
Fig.~\ref{fig:spectrev1-uarch-sem} contains two instances of the program's microarchitectural semantics deviating from what is architecturally implied---the two dashed \texttt{rf} edges are lacking consistent \texttt{rfx} edges. The endpoints of these culprit \texttt{com} edges---$\bot{}$ for both---constitute \textbf{\textit{receivers}} of microarchitectural leakage.
In this paper, we define three classes of \textbf{\textit{transmitters}} which can microarchitecturally convey information to a receiver, described as follows.

First, \textit{\xstate{} transmitters} are instructions which source (i.e., convey information to) a receiver via an \texttt{rfx} edge. In other words, \xstate{} transmitters communicate some function of their accessed \xstate{} to a receiver via microarchitectural information flows.  In this paper, where \xstate{} consists of core-private cache lines or store buffer state which facilitate microarchitectural communication between same-address memory accesses, \xstate{} transmitters are in reality \textbf{\textit{address transmitters}}---they transmit a function of their address operand.

Second, \textbf{\textit{data transmitters}} (resp. \textbf{\textit{control-flow transmitters}}) are address transmitters which are the target of an \texttt{addr} (resp. \texttt{ctrl}) dependency, originating at a \texttt{Read} $r$.
Both data transmitters and control transmitters leak a function of the data returned by $r$, where $r$ is referred to as the \textbf{\textit{access instruction}}. However, we consider data transmitters more dangerous since control transmitters leak the outcome of a branch condition involving $r$'s return value rather than the return value itself.

Third, \textbf{\textit{universal data transmitters}} (resp. \textbf{\textit{universal control-flow transmitters}}) are data transmitters (resp. control-flow transmitters) whose access instruction is the target of an \texttt{addr} dependency, originating at some \texttt{Read} $r'$. For example, a chain of the form $r'\xrightarrow[]{\texttt{addr}} access\xrightarrow[]{\texttt{addr/ctrl}} transmit\xrightarrow[]{\texttt{rfx}}receiver$ indicates that the \textit{memory location} supplied to the access instruction
is controlled by the data returned by $r'$. If an adversary can control the contents of the memory location referenced by $r'$, it can read arbitrary memory~\cite{heretostay:arxiv:2019}. 
In Fig.~\ref{fig:spectrev1-uarch-sem}, instructions 2, 5, and 6 are address transmitters, 5 and 6 are data transmitters, and 6 is a universal data transmitter.

\subsection{An Axiomatic Speculative Semantics}
\label{subsec:whyspeculation}

It is crucial that \lcms{} account for all instructions capable of accessing \xstate{}---and thus all instructions capable of impacting the \texttt{comx} relation---including those that \textit{transiently} execute.
Thus, \lcms{} extend MCMs with a \textit{speculative semantics}~\cite{oleksenko:specfuzz,guarnieri:spectector,ctfoundations:pldi:20}, as illustrated in Fig.~\ref{fig:spectrev1-spec-sem}.

The speculative semantics of \lcms{} leverages a new \textit{transient fetch order} (\texttt{tfo}) relation to construct a per-thread total order on all instructions that are \textit{fetched} from instruction memory. \texttt{po} is a subset of \texttt{tfo}, and  \textit{instructions ordered by \texttt{tfo} but not \texttt{po} are considered transient}; \texttt{po} relates committed instructions only. Transient instructions can interact with other transient or committed instructions via \xstate{} and ultimately construct new opportunities for program-level information leakage by impacting a program's microarchitectural semantics.

We consider two types of hardware speculation in this paper---\textbf{\textit{control-flow speculation}} and \textbf{\textit{address speculation}}. To model control-flow speculation, at each control-flow instruction where the architectural semantics considers both possible committed branch paths (i.e., both possible event structures), the speculative semantics additionally considers a window of speculative instructions along each branch path according to a user-defined speculation depth. In this way, formal analyses that leverage LCMs consider the worst-case attacker who can poison the prediction of any branch~\cite{guarnieri:spectector, guarnieri:contracts}. Fig.~\ref{fig:spectrev1-spec-sem} demonstrates this idea with a speculation depth of two. The left (i.e., taken) branch speculatively jumps to the end of the program ($\bot{_S}$) before rolling back speculation and executing the body of the branch. The right (i.e., not-taken) branch speculatively executes the body of the branch (5$_S$ and 6$_S$) before rolling back speculation and jumping to the end of the program.

To model address speculation, we consider two types---\textbf{\textit{store forwarding} }and \textbf{\textit{alias prediction}}. Both enable an architectural \texttt{Read} instruction to induce a window of speculation.
Furthermore, they \textit{relax} the placement of \texttt{rfx} edges, and thus the derived \texttt{frx} edges, in legal candidate executions.
Figs.~\ref{fig:spectrev4}~and~\ref{fig:spectre-psf} in \S\ref{sec:casestudy} give examples of data leakage which results from store forwarding and alias prediction, respectively.

Store forwarding permits reads to forward values from older stores in the reorder buffer (ROB) whose addresses have resolved and whose data is ready. Such a store must be the most-recent older store to the same address among stores whose addresses have been resolved. However, \textit{all} older stores need not resolve their addresses before forwarding can occur. Thus, while a load will always read from the \textit{correct address}, it may be forwarded \textit{stale data} speculatively.
Alias prediction permits a load to forward from a store with a potentially \textit{mismatching address}---i.e., a load may forward data from a store even if its address has not resolved.

\section{The \toolkit{} Toolkit}
\label{sec:vocab}
\lcm{} \textit{event structures} match those of MCMs (\S\ref{sec:event-structures}), while \lcm{} \textit{candidate executions} (\S\ref{sec:candidate-executions}) additionally include \texttt{tfo} and \texttt{comx} relations.
Moreover, \texttt{po}-derived relations in MCMs, such as \texttt{dep}, are derived from \texttt{tfo} in \lcms{}.

We mechanize the \lcm{} vocabulary in a toolkit built in Alloy~\cite{alloy}, called \toolkit{}, which we plan to open-source. \toolkit{} is akin to similar MCM frameworks---e.g., the \texttt{herd} simulator that takes as input an axiomatic MCM specification defined in the \texttt{.cat} domain specific language (DSL)~\cite{alglave:herd}. \toolkit{} supports the design and formal analysis of custom \lcm{} specifications using our axiomatic vocabulary. This section highlights some features of \toolkit{}.

\textbf{Beyond Memory, Control-Flow, and Fence Events:}
\label{sec:morethanmem}
In this paper, we focus on hardware leakage that results from memory systems optimizations. Thus, our case studies (\S\ref{sec:casestudy} and \S\ref{sec:results}) consider the same set of \texttt{Events} as MCMs---memory,  control flow, and fences. However, since microarchitectural leakage can result from \xstate{} interactions between arbitrary program instructions, \toolkit{} supports defining \lcms{} which feature any \texttt{Events} (i.e., instructions) of the designer's choosing.

\textbf{Beyond Architectural State:}
\label{sec:lcmloc}
In both MCMs and \lcms{}, \texttt{Locations} are represented symbolically, as is \xstate{} in \lcms{}. Namely,
it is not relevant for MCM or security analyses what concrete hardware state is accessed on behalf of a given instruction. Instead, it is important that we can identify when a \textit{pair of instructions} may access the \textit{same hardware state}---i.e., when a pair of instructions may be related via \texttt{com} or \texttt{comx}.

\textbf{Modeling Microarchitectural Data-Flow:}
\label{sec:comx-subrosa}
To define an \lcm{} for a microarchitecture, rules for constructing legal instantiations of \texttt{comx} must be established. To do this, one must first identify the \textit{conflict sets} of the microarchitecture in question---sets of instructions that are capable of interacting via a common set of \xstate{} elements.
At the finest granularity, conflict sets may be defined by identifying \textit{data-flow registers} within a microarchitecture---hardware state elements that may be written to by some instructions and read from by others, thereby facilitating microarchitectural data-flow.

For a given data-flow register, the sorts of instructions that may access it during their execution comprise its corresponding conflict set. Recent work shows that it is possible to identify such data-flow registers with the help of formal RTL property verifiers~\cite{cadence:jasper_gold}. Further, as shown in this paper, it is possible to merge conflict sets which facilitate the same sort of data-flow.

\textbf{Modeling Transient Execution:}
\label{sec:incorporating}
\toolkit{} supports custom user-defined \textit{speculation primitives}~\cite{microsoft:taxonomy,microsoft:ssb} which constrain the legal placement of \texttt{tfo} edges in a candidate execution. Instructions defined as speculation primitives are capable of inducing transient execution.
Note that if an instruction is transiently executed,
\toolkit{} assumes that it executes completely with respect to its updates on \xstate{}.
\section{Detecting Leakage in Real-World Examples}
\label{sec:casestudy}
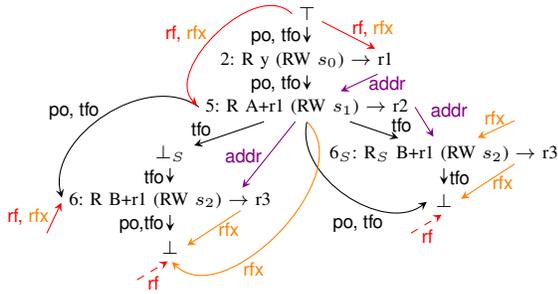
\begin{figure}[t]
        \begin{minipage}[b]{\linewidth}\vspace{-8pt}
        \centering
        \begin{tikzpicture}[->,>=stealth,shorten >=1pt,auto,node distance=.62cm, node/.style={rectangle,draw=none,fill=none,minimum size=1mm}]
            \scriptsize
            \node[node] (top) {$\top$};
            \node[node] (Ry) [below of=top] {2: R y (RW $s_0$) $\rightarrow$ r1};
            \node[node] (RA) [below of=Ry] {5: R A+r1 (RW $s_1$) $\rightarrow$ r2};
            \node[node] (space-left) [left of=RA, xshift=-1.2cm] {};
            \node[node] (space-right) [right of=RA, xshift=1.2cm] {};
            \node[node] (bottom-spec) [below of=space-left] {$\bot{_S}$};
            \node[node] (RB) [below of=bottom-spec] {6: R B+r1 (RW $s_2$) $\rightarrow$ r3};
            \node[node] (bottom-left) [below of=RB] {$\bot$};
            \node[node] (RB-spec) [below of=space-right] {6$_S$: R$_S$  B+r1 (RW $s_2$) $\rightarrow$ r3};
            \node[node] (bottom-right) [below of=RB-spec] {$\bot$};
            \coordinate[below right of=RB-spec, yshift=.3cm, xshift=.5cm] (RB-spec-bottom-right);
            \coordinate[below right of=RB, yshift=.3cm, xshift=.5cm] (RB-bottom-right);
            
            \coordinate[below left of=RB, xshift=-1.2cm] (RB-init);
            \coordinate[above right of=RB-spec, xshift=.5cm] (RB-spec-init);
            \coordinate[above right of=RA, xshift=.5cm] (RA-xinit);
            \coordinate[above right of=RB, yshift=-.3cm, xshift=.5cm] (RB-xinit);
            \coordinate[below left of=bottom-right] (bottom-right-obs);
            \coordinate[below left of=bottom-left] (bottom-left-obs);
            \coordinate[above right of=RA, xshift=.5cm] (RA-init);
            \coordinate[above right of=Ry, xshift=.5cm] (Ry-init);
            \coordinate[above left of=Ry, xshift=-.5cm] (Ry-xinit);
            \coordinate[below left of=RA, xshift=-.1cm] (RA-xinit);
            \coordinate[above right of=Ry, yshift=-.3cm, xshift=.5cm] (Ry-top-right);
            
            \path[every node/.style={font=\sffamily\scriptsize, fill=none,inner sep=2pt}]
                (top) edge [left] node {po, tfo} (Ry)
                (Ry) edge [left] node {po, tfo} (RA)
                (top) edge [red,  right, near start] node {~~rf, {\color{orange}rfx}} (Ry-top-right)
                (RA) edge [left] node[xshift=-4pt, yshift=2pt] {tfo} (bottom-spec)
                (RA) edge [right] node[xshift=4pt, yshift=2pt] {tfo} (RB-spec)
                (bottom-spec) edge [left] node {tfo} (RB)
                (RB) edge [left] node {po,tfo} (bottom-left)
                (RA.west) edge [bend right=70,left] node[yshift=2pt] {po, tfo} (RB.west)
                (RB-spec) edge [right] node {tfo} (bottom-right)
                (RA.south) edge [bend right=70, right, left, near end] node[yshift=-5pt] {po, tfo} (bottom-right.west)
                (RA) edge [violet, left] node {addr} (RB-xinit)
                (RA.east) edge [violet] node {addr} (RB-spec)
                (RA-init) edge [violet, right] node {~addr} (RA)
                (top.west) edge [red, left, bend right=90] node {rf, {\color{orange} rfx~}} (RA.west)
                (RA.south) edge [orange, right, bend left=90, near end] node {~~~rfx} (bottom-left.south)
                (RB-init) edge [red, left] node {~~rf, {\color{orange} rfx}} (RB.west)
                (RB-spec-init) edge [orange, above] node {rfx} (RB-spec)
                (bottom-right-obs) edge [red, below, dashed] node {rf} (bottom-right.south)
                (bottom-left-obs) edge [red, below, dashed] node {rf} (bottom-left.south)
                (RB-spec-bottom-right) edge [orange, right] node {rfx} (bottom-right.east)
                (RB-bottom-right) edge [orange, right] node {rfx} (bottom-left.east)
                ;
        \end{tikzpicture}\vspace{-10pt}
        \end{minipage}  
\caption{\textsc{Spectre v1} variant~\cite{jiyong:stt,guarnieri:contracts}. \lcms{} detect a transient transmitter and non-transient access. 
}
\label{fig:spectrev1-nonspec-access}
\end{figure}

In this section, we use \S\ref{sec:makingacaselcm}'s axiomatic \lcm{} vocabulary to formalize \textit{microarchitectural leakage}.
Our leakage definition is extensible, but focuses on formalizing hardware leakage on behalf of processor memory systems optimizations.



\subsection{Formalizing Microarchitectural Leakage}
\label{sec:detectingleaks}
To construct our leakage definition, we define mappings from the building blocks of an \lcms{}'s architectural semantics---the \texttt{rf}, \texttt{co}, and \texttt{fr}---to the building blocks of its microarchitectural semantics---the \texttt{rfx}, \texttt{cox}, and \texttt{frx}.
Our mappings assume that architectural memory events access a single \xstate{} location which collectively represents a core-local cache line and store buffer entry. While we could model these state elements as a pair of \xstate{} locations, we choose to merge them since they both facilitate microarchitectural data-flow between architectural memory instructions.
We also limit our mappings below to a \textit{single-core} execution setting where caches are \textit{direct-mapped} feature a \textit{write-allocate} cache policy.

If two writes are ordered by \texttt{co}, $w_0 \xrightarrow[]{co}w_1$, they should be similarly ordered by \texttt{cox} and \texttt{frx}. This is because two same-address writes behave as microarchitectural read-modify-writes (\S\ref{sec:xstate}) with respect to the same \xstate{}. Their writes to the store buffer and cache will be ordered, and $w_0$'s cache read will precede $w_1$'s cache write.
If $w_0$ immediately precedes $w_1$ in \texttt{co}, the two writes should also be ordered by \texttt{rfx} in the absence of interference---i.e., $w_1$ should be microarchitecturally-sourced by $w_0$'s cache line (a cache hit).
As explained in \S\ref{sec:modeling-leakage}, if a write and read are ordered by \texttt{rf}, $w\xrightarrow[]{rf}r$, they should be similarly ordered by \texttt{rfx} in the absence of interference.
If a read and a write are ordered by \texttt{fr}, $r\xrightarrow[]{fr}w$, 
then they should be similarly ordered by \texttt{frx}---$r$ will microarchitecturally  read its cache line or store buffer entry before $w$ microarchitecturally writes.
\textit{A microarchitectural leak is detected when an architecturally-implied microarchitectural relation is missing from a consistent candidate execution.}

\subsection{\lcms{} by Example}
\label{sec:specexamples}
We show that the \lcm{} vocabulary faithfully detects leakage in a sampling of (transient and non-transient) microarchitectural attacks form the literature. In all examples, dashed edges denote \texttt{com} relations with \texttt{comx} inconsistencies---they ``point to'' (via a directed edge) receivers, which are used to identify transmitters according to the rules in \S\ref{sec:modeling-leakage}. All numerical instruction identifiers refer to candidate execution graphs. Moreover, some edges are omitted in figures for clarity when they are not central to the exemplified leakage.
\textbf{\textsc{Spectre v1}:}
Fig.~\ref{fig:spectrev1-spec-sem} summarizes the candidate executions of vanilla \textsc{Spectre~v1}~\cite{spectre} (Fig.~\ref{fig:spectre-v1-C}). 
The program features a speculation primitive---a conditional branch which induces a window of speculation in each event structure instantiated by the branch (i.e, each fork of the graph). Two inconsistent \texttt{rf} edges point towards receivers---both $\bot$ nodes. Instruction 2 is an address transmitter; 5 and 5$_S$ are data transmitters with access instruction 2; 6 and 6$_S$ are universal data transmitters with access instructions 5 and 5$_S$, respectively.
Notably, some transmitters are transient while others are non-transient.

Fig.~\ref{fig:spectrev1-nonspec-access} shows another variant of \textsc{Spectre v1}~\cite{jiyong:stt, guarnieri:contracts} (code below) featuring the same speculation primitive---a conditional branch---and the same universal data transmitters---6 and 6$_S$.
\begin{lstlisting}
x = A[y];
if(y < size_A)
  temp &= B[x];
\end{lstlisting}
However, this time the access instruction corresponding to both universal data transmitters (instruction 5) is non-transient. In Fig~\ref{fig:spectrev1-spec-sem}, the access instruction corresponding to instruction 6 is non-transient (instruction 5) while the access instruction corresponding to instruction 6$_S$ is transient (instruction 5$_S$).

Notably, STT~\cite{jiyong:stt} declared preventing the leakage of non-transiently accessed data as out of scope, although other related work captures leakage of this sort~\cite{guarnieri:spectector,ctfoundations:pldi:20}. 

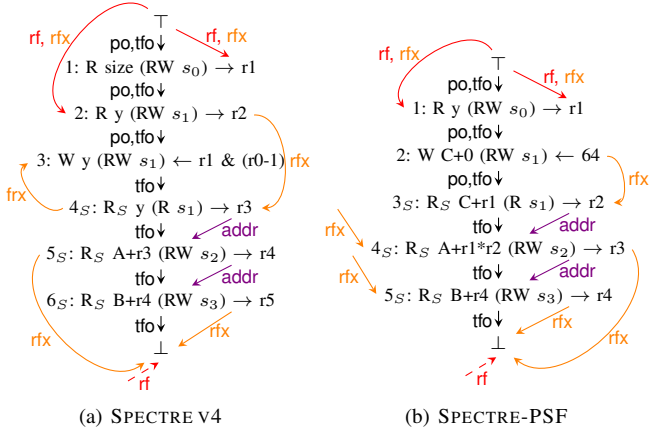
\begin{figure}[t]
\subfloat[\textsc{Spectre v4}]{
        \label{fig:spectrev4}
\begin{minipage}[b]{.45\linewidth}
    \centering
        \begin{tikzpicture}[->,>=stealth,shorten >=1pt,auto,node distance=.62cm, node/.style={rectangle,draw=none,fill=none,minimum size=1mm}]
            \scriptsize
            \node[node] (top) {$\top$};
            \node[node] (Rsize) [below of=top] {1: R size (RW $s_0$) $\rightarrow$ r1};
            \node[node] (Ry0) [below of=Rsize] {2: R y (RW $s_1$) $\rightarrow$ r2};
            \node[node] (Wy) [below of=Ry0] {3: W y (RW $s_1$) $\leftarrow$ r1 \& (r0-1)};
            \node[node] (Ry1) [below of=Wy] {4$_S$: R$_S$ y (R $s_1$) $\rightarrow$ r3};
            \node[node] (RA) [below of=Ry1] {5$_S$: R$_S$ A+r3 (RW $s_2$) $\rightarrow$ r4};
            \node[node] (RB) [below of=RA] {6$_S$: R$_S$ B+r4 (RW $s_3$) $\rightarrow$ r5};
            \node[node] (bottom) [below of=RB] {$\bot$};
            
            \coordinate[below left of=bottom] (bottom-obs);
            \coordinate[above right of=Rsize, xshift=.5cm] (Rsize-init);
            \coordinate[above right of=Ry0, xshift=.5cm] (Ry-init);
            \coordinate[above right of=Rsize, yshift=-.3cm, xshift=.5cm] (Rsize-top-right);
            \coordinate[below right of=RB, yshift=.3cm, xshift=.5cm] (RB-bottom-right);
            \coordinate[below right of=RA, yshift=.3cm, xshift=.5cm] (RA-bottom-right);
            \coordinate[below right of=Ry1, yshift=.3cm, xshift=.5cm] (Ry1-bottom-right);
            
            \path[every node/.style={font=\sffamily\scriptsize, fill=none,inner sep=2pt}]
                (top) edge [left] node {po,tfo} (Rsize)
                (Rsize) edge [left] node {po,tfo} (Ry0)
                (Ry0) edge [left] node {po,tfo} (Wy)
                (Wy) edge [left] node {tfo} (Ry1)
                (Ry1) edge [left] node {tfo} (RA)
                (RA) edge [left] node {tfo} (RB)
                (RB) edge [left] node {tfo} (bottom)
                (bottom-obs) edge [red, below, dashed] node {rf} (bottom.south)
                (Ry1) edge [orange, bend left=90, left] node {frx} (Wy)
                (top) edge [red, bend right=90, left] node {rf, {\color{orange}rfx}} (Ry0.west)
                (Ry0) edge [orange, bend left=90, right] node {rfx} (Ry1)
                (RA.west) edge [orange, bend right=80, left] node {rfx~~} (bottom)
                (top) edge [red,  right, near start] node {~~rf, {\color{orange}rfx}} (Rsize-top-right)
                (RB-bottom-right) edge [orange, right] node {rfx} (bottom.east)
                (RA-bottom-right) edge [violet, right] node {~addr} (RB)
                (Ry1-bottom-right) edge [violet, right] node {~addr} (RA)
                ;
        \end{tikzpicture}
\end{minipage} }
\subfloat[\textsc{Spectre-PSF}]{
        \label{fig:spectre-psf}\begin{minipage}[b]{.5\linewidth}
    \centering
        \begin{tikzpicture}[->,>=stealth,shorten >=1pt,auto,node distance=.62cm, node/.style={rectangle,draw=none,fill=none,minimum size=1mm}]
            \scriptsize
            \node[node] (top) {$\top$};
            \node[node] (Ry) [below of=top] {1: R y (RW $s_0$) $\rightarrow$ r1};
            \node[node] (WC0) [below of=Ry] {2: W C+0 (RW $s_1$) $\leftarrow$ 64};
            \node[node] (RC) [below of=WC0] {3$_S$: R$_S$ C+r1 (R $s_1$) $\rightarrow$ r2};
            \node[node] (RA) [below of=RC] {4$_S$: R$_S$ A+r1*r2 (RW $s_2$) $\rightarrow$ r3};
            \node[node] (RB) [below of=RA] {5$_S$: R$_S$ B+r4 (RW $s_3$) $\rightarrow$ r4};
            \node[node] (bottom) [below of=RB] {$\bot$};
            
            \coordinate[below left of=bottom] (bottom-obs);
            \coordinate[above right of=Ry, xshift=.5cm] (Ry-init);
            \coordinate[below left of=bottom] (bottom-obs);
            \coordinate[below right of=RC, yshift=.3cm, xshift=.5cm] (RC-bottom-right);
            \coordinate[below right of=RA, yshift=.3cm, xshift=.5cm] (RA-bottom-right);
            \coordinate[below right of=RB, yshift=.3cm, xshift=.5cm] (RB-bottom-right);
            \coordinate[above right of=Rsize, yshift=-.3cm, xshift=.5cm] (Ry-top-right);
            \coordinate[above left of=RA, yshift=.1cm, xshift=-1.7cm] (RA-top-left);
            \coordinate[above left of=RB, yshift=.1cm, xshift=-1.5cm] (RB-top-left);
            
            \path[every node/.style={font=\sffamily\scriptsize, fill=none,inner sep=2pt}]
          	    (top) edge [red,  right, near start] node {~~rf, {\color{orange}rfx}} (Ry-top-right)
                (top) edge [left] node {po,tfo} (Ry)
                (top) edge [red, bend right=90, left] node {rf, {\color{orange}rfx}~~} (Ry.west)
                (Ry) edge [left] node {po,tfo} (WC0)
                (WC0) edge [left] node {po,tfo} (RC)
                (RC) edge [left] node {tfo} (RA)
                (RA) edge [left] node {tfo} (RB)
                (RB) edge [left] node {tfo} (bottom)
                (RB-top-left) edge [left, orange] node {rfx} (RB.west)
                (RA-top-left) edge [left, orange] node {rfx} (RA.west)
                (WC0) edge [orange, bend left=90, right] node {rfx} (RC)
                (RA.east) edge [orange, bend left=80, right] node {rfx~} (bottom)
                (bottom-obs) edge [red, below, dashed] node {rf} (bottom.south)
                (RC-bottom-right) edge [violet, right] node {~addr} (RA)
                (RA-bottom-right) edge [violet, right] node {~addr} (RB)
                (RB-bottom-right) edge [orange, right] node {~rfx} (bottom)
                ;
        \end{tikzpicture}
\end{minipage} }
\caption{\textsc{Spectre v4}~\cite{spectrev4,spectrev3av4} and \textsc{Spectre-PSF}~\cite{guanciale:inspectre,ctfoundations:pldi:20}.
\lcms{} detect a transient transmitter and transient access. 
}

\label{fig:spectrev4-psf}
\end{figure}

\textbf{\textsc{Spectre v4}:}
Fig.~\ref{fig:spectrev4} is representative of \textsc{Spectre~v4}, described by the code below.
\begin{lstlisting}
y = y & (size_A - 1);
x = A[y];
temp &= B[x];
\end{lstlisting}
The speculation primitive is \textit{store forwarding}
(\S\ref{subsec:whyspeculation})---instruction 4 reads from a stale same-address
write. 
The \texttt{frx} relation between instructions 4$_S$ and 3 illustrates this behavior.
\texttt{frx} edges can also be understood as \textit{reads-before}. In other words, 4$_S$ reads from \xstate{} element s$_1$ \textit{before} s$_1$ is overwritten by 3. The figure also illustrates that instruction 4 is microarchitecturally sourced from the first read of \texttt{y}, namely instruction 2, via an \texttt{rfx} relation. Ultimately, this behavior leads to 6$_S$ manifesting as a transient universal data transmitter of data transiently accessed by instruction $5_S$. Also, 5$_S$ is a transient data transmitter, with transient access instruction 4$_S$.

We note that \textsc{Spectre~v4} exhibits particularly interesting microarchitectural behavior that is relevant for developing \lcms{} for Intel x86 microarchitectures (given that \textsc{Spectre~v4} has been observed on Intel processors~\cite{spectrev4, spectrev3av4}).
In particular, formally specifying an \lcm{} for a particular ISA requires defining a \textit{confidentiality predicate}
(\S\ref{sec:comx}). Consider the consistency predicate for TSO from \S\ref{sec:consistency-predicates} which is
the conjunction of the \textit{sc\_per\_loc}, \textit{rmw\_atomicity}, and \textit{causality} auxiliary predicates. Naively lifting \textit{sc\_per\_loc}
to constrain \texttt{comx} results in \textit{sc\_per\_loc\_x}~= {\tt acyclic(\{rfx~+~cox~+~frx~+~tfo\_loc\})}, where \texttt{tfo\_loc} is defined as \texttt{po\_loc} by substituting \texttt{tfo} for \texttt{po}.
This straightforward predicate derivation would rule out the execution in Fig.~\ref{fig:spectrev4}, which is in fact \textit{possible} on x86 microarchitectures. An \lcm{}, for Intel x86 processors (which permits~\textsc{Spectre v4}) must clearly permit cycles in \texttt{frx + tfo\_loc} in its confidentiality predicate.

\textbf{\textsc{Spectre-PSF}:}
Fig.~\ref{fig:spectre-psf} features a variant of \textsc{Spectre v4}~\cite{ctfoundations:pldi:20,guanciale:inspectre}, coined \textsc{Spectre-PSF}~\cite{cat-spectre} (code listing below).
\begin{lstlisting}
uint8_t A [16];
uint8_t C[2] = {0, 0};
if (y < size_C)
  C [0] = 64;
  temp &= B[A[C[y] * y]];
\end{lstlisting}
The speculation primitive is \textit{alias prediction}.
In particular, instruction 3 reads from an incorrect memory location---illustrated by the \texttt{rfx} edge between instructions 2 and 3.
This behavior leads to a transient universal data transmitter (5$_S$) with a transient access instruction (4$_S$).

\textsc{Spectre-PSF} also features interesting execution behavior that can influence the placement of \texttt{rfx} edges in \lcm{} candidate executions. Namely, read instructions can mis-predict the \texttt{xstate} they access such that they can
microarchitecturally read data written by prior stores to different addresses.

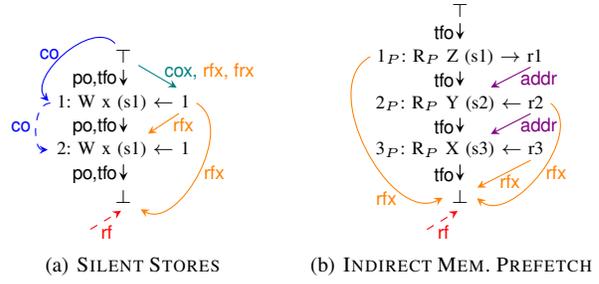
\begin{figure}[t]
\subfloat[\textsc{Silent Stores}]{
        \label{fig:silentstores}
        \begin{minipage}[b]{.42\linewidth}
    \centering
        \begin{tikzpicture}[->,>=stealth,shorten >=1pt,auto,node distance=.62cm, node/.style={rectangle,draw=none,fill=none,minimum size=1mm}]
            \scriptsize
            \node[node] (top) {$\top$};
            \node[node] (Wx0) [below of=top] {1: W x (s1) $\leftarrow$ 1};
            \node[node] (Wx1) [below of=Wx0] {2: W x (s1) $\leftarrow$ 1};
            \node[node] (bottom) [below of=Wx1] {$\bot$};
            
            \coordinate[below left of=bottom] (bottom-obs);
            \coordinate[below right of=Wx0, yshift=.3cm, xshift=.3cm] (Wx0-bottom-right);
            \coordinate[above right of=Wx0, yshift=-.25cm, xshift=.3cm] (Wx0-top-right);

            \path[every node/.style={font=\sffamily\scriptsize, fill=none,inner sep=2pt}]
                (top) edge [left] node {po,tfo} (Wx0)
                (Wx0) edge [left] node {po,tfo} (Wx1)
                (Wx1) edge [left] node {po,tfo} (bottom)
                (top) edge [blue, bend right=90, left] node {co} (Wx0.west)
                (Wx0.west) edge [blue, bend right=90, left, dashed] node {co} (Wx1.west)
                (top) edge [teal,  right, near start] node {~~cox, {\color{orange}rfx, frx}} (Wx0-top-right)
                (Wx0.east) edge [orange, bend left=90, right] node {rfx} (bottom)
                (bottom-obs) edge [red, below, dashed] node {rf} (bottom.south)
                (Wx0-bottom-right) edge [orange, right] node {~rfx} (Wx1)
                ;
        \end{tikzpicture}
        \end{minipage} 
        }
\subfloat[\textsc{Indirect Mem. Prefetch}]{
        \label{fig:imp}
        \begin{minipage}[b]{.48\linewidth}
    \centering
        \begin{tikzpicture}[->,>=stealth,shorten >=1pt,auto,node distance=.62cm, node/.style={rectangle,draw=none,fill=none,minimum size=1mm}]
            \scriptsize
            \node[node] (top) {$\top$};
            \node[node] (RZ) [below of=top] {1$_P$: R$_P$ Z (s1) $\rightarrow$ r1};
            \node[node] (RY) [below of=RZ] {2$_P$: R$_P$ Y (s2) $\leftarrow$ r2};
            \node[node] (RX) [below of=RY] {3$_P$: R$_P$ X (s3) $\leftarrow$ r3};
            \node[node] (bottom) [below of=RX] {$\bot$};
            
            \coordinate[below left of=bottom] (bottom-obs);
            \coordinate[above right of=RX, xshift=.5cm] (RX-xinit);
            \coordinate[above right of=RY, xshift=.5cm] (RY-xinit);
            \coordinate[above right of=RZ, xshift=.5cm] (RZ-xinit);
            \coordinate[below right of=RZ, yshift=.3cm, xshift=.5cm] (RZ-bottom-right);
            \coordinate[below right of=RY, yshift=.3cm, xshift=.5cm] (RY-bottom-right);
             \coordinate[below right of=RX, yshift=.3cm, xshift=.5cm] (RX-bottom-right);   
            
            \path[every node/.style={font=\sffamily\scriptsize, fill=none,inner sep=2pt}]
                (top) edge [left] node {tfo} (RZ)
                (RZ) edge [left] node {tfo} (RY)
                (RY) edge [left] node {tfo} (RX)
                (RX) edge [left] node {tfo} (bottom)
                (RY.east) edge [orange, bend left=90, right] node {rfx} (bottom.east)
                (RZ.west) edge [orange, bend right=90, left, near end] node[yshift=-.1cm] {rfx} (bottom.west)
                (bottom-obs) edge [red, below, dashed] node {rf} (bottom.south)
                (RZ-bottom-right) edge [violet, right] node {~addr} (RY)
                (RY-bottom-right) edge [violet, right] node {~addr} (RX)
                (RX-bottom-right) edge [orange, right, near end] node {~rfx} (bottom)
                ;
        \end{tikzpicture}
\end{minipage} 
}
\caption{\textsc{Non-Spectre}~\cite{pandora:isca:21}. In \protect\subref{fig:silentstores}, \lcms{} detect a non-transient transmitter of a non-transiently accessed \xstate{}. In \protect\subref{fig:imp}, 
IMPs can construct a universal data transmitter of 
prefetched data.
}
\label{fig:non-spectre}
\end{figure}

\textbf{Non-Spectre Attacks:}
Recent work shows that various microarchitectural optimizations can be leveraged to leak program data in a manner as severe as Spectre attacks~\cite{pandora:isca:21}. Fig.~\ref{fig:non-spectre} features programs that exercise two such optimizations.
Fig.~\ref{fig:silentstores} features leakage on hardware that implements \textit{silent stores}~\cite{silentstores}, which avoid explicitly writing to memory when a store's data operand matches the current contents at its effective address. Here, instruction 2 is an \xstate{} transmitter. Unlike most \xstate{} transmitters in this paper, instruction 2 transmits the data field of its accessed \xstate{} $s_0$ rather than the address field.
This is because the silent store optimization triggers based on the result of a \textit{data comparison} while a cache hit/miss is triggered based on the result of an \textit{address comparison}.

Fig.~\ref{fig:imp} features leakage on hardware implementing an \textit{indirect memory prefetcher} (IMP)~\cite{yu:imp} (recently patented by Intel~\cite{intel-imp}). Hardware featuring such a prefetcher tries to detect programs of the form \textit{for(i = 0...N) X[Y[Z[i]]]} and prefetch the cache line corresponding to \textit{\&X[Y[Z[i + $\Delta$]]]}.
The security implications of this optimization are discussed in recent work~\cite{pandora:isca:21}.
Notably, the authors point out an IMP can construct a \textit{universal read gadget}~\cite{heretostay:arxiv:2019}, and Fig.~\ref{fig:imp} indeed indicates that prefetch instruction $3_P$ is a universal data transmitter.
\section{\tool{}: Detecting Leakage with \lcms{}}
\label{sec:results}
We develop a static analysis tool, \tool{}, based on \lcms{} which automatically identifies and repairs Spectre vulnerabilities in programs.
Our approach is inspired by a tool which restores sequentially consistency, via automated fence insertion, for programs running on hardware implementing weak MCMs~\cite{alglave:dsotf}.
\tool{} is implemented as a custom IR pass in LLVM~\cite{LattnerAdve:llvm}. It takes a \textit{\textbf{C source file as input}}, compiles it to LLVM IR using \textsc{Clang} v12.0.0, and analyzes each defined function one-by-one.
Eventually, \tool{} \textit{\textbf{outputs a list of transmitters and a set of consistent candidate executions}} (in graph form) which give witness to detected software vulnerabilities. \tool{} can also \textit{\textbf{automatically insert mitigations}} (e.g., fences, like Intel's \texttt{lfence}) to repair vulnerable programs.

\subsection{Constructing an Abstract CFG (A-CFG)}
\tool{} first transforms a function's
LLVM IR control-flow graph (CFG) into a loop/call-free \textit{Abstract CFG (A-CFG)}---our name for a CFG that has undergone loop/function summarizaiton and function inlining.

\textbf{Loop Summarization:}
\label{sec:loop-summaries}
To eliminate a loop from a function's CFG, \tool{} \textit{summarizes} all of the ways in which it could be involved in hardware-induced leakage using a finite (and minimal) number of instructions as follows.
First, recall that \lcms{} detect microarchitectural leakage by
comparing architecture-level (via \texttt{com}) and microarchitecture-level (via \texttt{comx}) instruction interactions. 
This suggests a loop summarization approach which accounts for (1) how instructions in any loop instance can interact with instructions outside of the loop, and (2) how instructions in two arbitrary loop instances can interact with each other. 
Second, consider a memory alias analysis procedure (\S\ref{sec:alias-analysis}) that can summarize for all memory accesses in the loop the set of virtual memory locations it may access across all iterations.
We conclude that with memory alias analysis, all relevant \texttt{com}/\texttt{comx} interactions involving loop instructions can be modeled with just two loop un-rollings.



\textbf{Function Inlining:}
With loops summarized, \tool{} inlines all function calls.
Recursive calls are inlined twice via similar logic to that which enables loop summarization.
For a call whose target function is not defined, \tool{} interprets it as a load \textit{or} store to one of its pointer operands---e.g., 
\texttt{memcmp(void *dst, const void *src, size\_t n)}
can behave as a load or store to \texttt{*dst} or \texttt{*src}. 
An SMT solver
considers all possible options when searching for a way to construct a candidate execution featuring leakage.




\subsection{Constructing a Symbolic Abstract Event Graph (S-AEG)}
\tool{} extends an A-CFG  to produce a \textit{Symbolic Abstract Event Graph (S-AEG)}---an over-approximation of all of the corresponding function's possible candidate executions.\footnote{
Compared to AEGs in prior work~\cite{alglave:dsotf}, S-AEGs are encoded more compactly as a set of first-order logic formulas, rather than as explicit graph data structures.
}
An S-AEG
features exactly the same set of nodes as the A-CFG from which it is derived. 
However, four categories of symbolic edges are added:
control-flow (\texttt{po} and \texttt{tfo}), \texttt{dep}, \texttt{com}, and \texttt{comx}.
Moreover, symbolic variables are associated with each S-AEG node and edge.
Legal assignments to these variables are constrained by a set of first-order logic formulas that describe what constitutes a consistent candidate execution---including consistency and confidentiality predicates. Deriving a concrete candidate execution from an S-AEG may then be achieved by searching for a variable assignment which satisfies said formulas.
Other than the assumption the target hardware features write-allocate caches and does not implement silent stores~\cite{silentstores}, we conservatively leave \texttt{comx} unconstrained; \texttt{com} is constrained by the TSO consistency predicate (\S\ref{sec:consistency-predicates}).

\textbf{Design decisions:} We currently make a few key design decisions regarding S-AEG construction. First, we assume \lcms{} where only memory instructions induce \xstate{} accesses, under the assumptions outlined in \S\ref{sec:xstate}.
Second, we assume a one-to-one correspondence between architectural addresses and modeled \xstate{} locations.
This assumption may result in \tool{} uncovering leakage that would be impossible due to cache collisions which are realized for a particular cache architecture.
Third, given our empirical observations, \tool{} also considers a special type of \texttt{addr},
called \texttt{addr\_gep} (\textit{get-element-pointer address dependency}). \texttt{addr\_gep} maps a \texttt{Read} to a \texttt{MemoryEvent}, where the \texttt{Read}'s return value is added to a base address to compute the \texttt{MemoryEvent}'s effective address.\footnote{LLVM IR features a pointer-only arithmetic instruction, called \texttt{getelementptr}, which signifies such behavior.}
Distinguishing \texttt{addr\_gep} from other \texttt{addr} dependencies---which indicate the source instruction supplies a base address---enables \tool{} to filter out benign leaks (\S\ref{sec:libsodium}).

\textbf{Alias Analysis:} \label{sec:alias-analysis}
\tool{} uses an alias analysis procedure to 
reduce the search space when looking for transmitters.
First, \tool{} selectively applies LLVM's built-in alias analysis~\cite{LattnerAdve:llvm} to the S-AEG, only including constraints (that assert particular address pairs are unequal) when they are valid under \tool{}'s CFG-to-A-CFG transformation.
Next, \tool{} assumes that (1) all S-AEG stack allocations have distinct addresses and
(2) alias analysis results hold during transient execution. These assumptions
do not restrict the set of discoverable transmitters.

\subsection{Leakage Detection Engines}
\label{sec:leakage-engines}
Once an S-AEG has been constructed, \tool{} is ready to search the graph for potential transmitters. \tool{} initiates this procedure by all adding constraints encoded in the S-AEG to a Z3 SMT solver instance \cite{moura:z3}. These constraints define the space of consistent candidate executions of the function under evaluation.
The next intuitive step is to directly encode as a constraint the expected behaviors of a leakage-free program (according to \S\ref{sec:detectingleaks}), so that Z3 can search for violations of this constraint.
Unsurprisingly, this approach produces a \textit{large} number of transmitters, all of which are not equally interesting/dangerous.
Thus, we develop multiple optimized leakage detection engines that perform a directed search for specific types of leakage, parameterized by the types of transmitters they are searching for. In this paper, we build leakage detection engines for \textsc{Spectre v1} and \textsc{Spectre v4}. 
\S\ref{sec:casestudy} shows that Spectre attacks violate the \texttt{rf} condition of our leakage definition in \S\ref{sec:detectingleaks}. Thus, \tool{}'s \textsc{Spectre v1} and \textsc{Spectre v4} detection engines both directly look for candidate executions which violate of this condition---the result is a set of candidate transmitters. \tool{} iterates over the candidate transmitters to find those which are also data/control transmitters or universal transmitters according to the user's preference. Per
\S\ref{sec:modeling-leakage}, a data/control transmitter manifests as a $a\xrightarrow[]{\texttt{addr/ctrl}} t\xrightarrow[]{\texttt{rfx}}r$ code pattern, and a universal data/control transmitter manifests as $r'\xrightarrow[]{\texttt{addr}} a\xrightarrow[]{\texttt{addr/ctrl}} t\xrightarrow[]{\texttt{rfx}}r$.
In reality, an \texttt{addr} edge in this definition can be realized as an \texttt{addr} edge followed by zero or more \texttt{data.rf} edges---i.e., \texttt{addr.(data.rf)$^*$}.
This means the value returned by an \texttt{addr}-dependent \texttt{Read} can be stored (\texttt{data}) and re-loaded (\texttt{rf}) any number of times before use.

\tool{}'s \textsc{Spectre v1} and \textsc{Spectre v4} detection engines differ based on the speculation primitives they consider---\textit{control-flow speculation} versus \textit{store forwarding}, respectively.






\subsection{Analyzing Spectre Benchmarks with \tool{}} 
\label{sec:benchmarks}
\begin{table}[t]
    \vspace{7pt}
    \centering
    \setlength{\tabcolsep}{1.5pt}
    \begin{tabular}{c|c|c|c|c|c}
    {\it \textsc{Spectre v1}} & {\it Intended} & {\it Detected} & {\it \textsc{Spectre v4}} & {\it Intended} & {\it Detected} \\
    {\it Benchmarks} & {\it Leakage} & {\it Leakage} & {\it Benchmarks} & {\it Leakage} & {\it Leakage} \\ \hline
     PHT01 & $U_D$ & $U_D$ & STL01      & $D$      & $D, \mathbf{U_D}$ \\\hline
     PHT02 & $U_D$ & $U_D$ & STL02      & $U_D$    & $U_D$    \\\hline
     PHT03 & $U_D$ & $U_D$ & STL03      & (none)   & $(U_D)^1$ \\\hline
     PHT04 & $U_D$ & $U_D$ & STL04      & $D$      & $D, (U_D)^1$ \\\hline
     PHT05 & $U_D$ & $U_D$ & STL05      & $U_D$    & $U_D$    \\\hline
     PHT06 & $U_D$ & $U_D$ & STL06      & $U_D$    & $U_D$    \\\hline
     PHT07 & $U_D$ & $U_D$ & STL07      & $U_D$    & $U_D$    \\\hline
     PHT08 & $U_D$ & $U_D$ & STL08      & $U_D$    & $U_D$    \\\hline
     PHT09 & $U_D$ & $U_D$ & STL09      & (none)   & $(D)^1, (U_D)^{1,2}$ \\\hline
     PHT10 & $U_C$ & $U_C$ & STL09\_bis & $D$      & $D, (U_D)^1$ \\\hline
     PHT11 & $U_D$ & $U_D$ & STL10      & $U_D$    & $U_D$    \\\hline
     PHT12 & $U_D$ & $U_D$ & STL11      & $D$      & $D, (U_D)^1$ \\\hline
     PHT13 & $U_D$ & $U_D$ & STL12      & (none)   & $(U_D)^1$    \\\hline
     PHT14 & $U_D$ & $U_D$ & STL13      & $D^*$    & $\mathbf{D}, (U_D)^1$ \\\hline
     PHT15 & $U_D$ & $U_D$ & ---        & ---      & --- \\\hline
    \end{tabular}
    \vspace{7pt}
    \caption{
    \tool{}'s evaluation of \textsc{Spectre v1}~\cite{kocher:sepcterv1-benchmarks} and \textsc{Spectre v4}~\cite{spectrev4-benchmarks} benchmarks.
    \textbf{Bold} = newly discovered leakage;
    $(\cdot)^1$ = false positive due semantic analysis imprecision;
    $(\cdot)^2$ = false positive due to imprecise loop summarization.
    }
    \label{tab:benchmark-eval}
\end{table}

We use \tool{} to analyze 15 \textsc{Spectre v1} (\textbf{PHT})~\cite{kocher:sepcterv1-benchmarks} and 14 \textsc{Spectre v4} (\textbf{STL})~\cite{spectrev4-benchmarks} benchmark programs.
Table~\ref{tab:benchmark-eval} shows our results, with each benchmark analyzed in \textit{less than half a second} on average.
A speculation depth ($d_\text{spec}$) of 250 (approximating ROB size) is used,
but not nearly exhausted.
We run \tool{} on each program and record the type(s) of transmitters it detects (\textit{Detected Leakage}).
We also manually inspect each program and record the
type of transmitter it is intended to feature (\textit{Intended Leakage}), using annotations from the benchmark authors. We find three types: data (\textit{D}), universal data ($U_D$), and  universal control ($U_C$).


When analyzing the \textbf{PHT} programs, \tool{} identifies all intended transmitters and constructs candidate execution graphs as witnesses. 
\tool{} also identifies a \textit{new attack variant} in all PHT programs---a data transmitter involving a transient instruction prefetching a cache line for a non-transient \textit{\texttt{tfo}-prior} instruction. \textit{Speculative interference attacks}~\cite{behnia:specinterfere} exhibit a similar phenomenon. We omit the finding from Table~\ref{tab:benchmark-eval} since the data transmitter is less dangerous than all \textit{intended} leakage.

For the \textbf{STL} programs,
compared with the benchmark authors' apparent intention,
\tool{} identifies \textit{more transmitters} that are in some cases \textit{more severe}.
In STL01 (below), for example, \tool{} identifies the intended transmitter---a data transmitter. However, it identifies {\it higher-severity} leakage that promotes said data transmitter to a {\it universal} data transmitter.
\begin{lstlisting}
void case_1(uint32_t idx) {
  uint32_t ridx=idx&(ary_size-1); // universal
  uint8_t **pp=&sec_ary; uint8_t ***ppp=&pp;
  (**ppp)[ridx]=0; // data 
  tmp &= pub_ary[sec_ary[ridx]];} // transmitter
\end{lstlisting}
STL01 intends to show that the access to \texttt{sec\_ary[ridx]} in line 5  can transiently read stale data before it is overwritten in line 4, rendering the access to access to \texttt{pub\_array} a data transmitter.
However, \tool{} finds a candidate execution where the same transmitter facilitates \textit{universal} data leakage---line 5's access to \texttt{idx} can read stale data before it is overwritten in line 2.
\tool{} also finds that STL13 is \textit{erroneously labeled as ``secure''} in the benchmark \textit{and} flagged as secure by the benchmark authors' formal tool~\cite{2021:huntingthehaunter} ---it features data leakage when a return instruction bypasses a store to the stack.


\tool{} detects some false positive leakage when analyzing \textbf{STL} programs due to two sources of imprecision. First, \tool{} does not perform semantic analysis of instructions. Thus, it cannot reason about the implications of index masking, a mitigation technique used by many STL programs.
Also, \tool{} does not consider the impact of loops on speculation depth when summarizing them. Thus, false positive leakage involving instructions which cannot exist in the processor simultaneously (due to ROB size) may be flagged (e.g., in STL09). 

STL03 and STL12 are intended to be \textit{safe} due to their use of C's \texttt{register} keyword to prevent \textit{storing} an array index.
We find that \textsc{Clang} \texttt{-O0} disregards the \texttt{register} keyword and stores the index in memory anyway, enabling it to be bypassed. 
Thus, we manually repair the LLVM IR output to create the
effect of \texttt{register}.
Table~\ref{tab:benchmark-eval}'s results incorporate this fix. 
One \textit{other} instance of false positive leakage is flagged, but the use of \texttt{register} repairs the intended leakage. 

A notable feature of \tool{} is its ability to insert a minimal number of fences to repair \textsc{Spectre v1} and \textsc{Spectre v4} leaks. We direct \tool{} to perform fence insertion in the \textbf{PHT} and \textbf{STL} benchmarks and confirm that all initially-detected leakage is mitigated with one fence per vulnerable program.



\subsection {Analyzing \libsodium{} with \tool{}}
\label{sec:libsodium}
\begin{figure}
    \centering
  \begin{tikzpicture}
  \begin{axis}[
    xmode=log,
    ymode=log,
    xlabel={\footnotesize\textit{S-AEG function size (node count)}},
    ylabel={\footnotesize\textit{runtime (s)}},
    width=8.5cm,
    height=3.8cm,
    log ticks with fixed point,
    x tick label style={
		/pgf/number format/.cd,
        use comma,
        1000 sep={}},
	ticklabel style={font=\footnotesize},
    legend pos=south east,
  ]
    \addplot[
        only marks,
        mark size=0.5pt,
        color=blue,
        mark=*,
    ] table[] {results/v1-ser.csv};
    \addlegendentry{\scriptsize\textsc{Spectre v1}}

    
    \addplot[
        only marks,
        mark size=0.5pt,
        color=red,
        mark=*,
    ] table[] {results/v4-ser.csv};
    \addlegendentry{\scriptsize\textsc{Spectre v4}}
    
  \end{axis}
  \end{tikzpicture}

    \caption{
    Serial CPU runtime vs. function size for \tool{}'s \libsodium{} analysis with ($d_\text{spec}$, $W_\text{size}$) set to (250, 500)/(25, 50) for \textsc{Spectre v1/v4}. No functions time out.
    }
    \label{fig:libsodium-perf}
\end{figure}
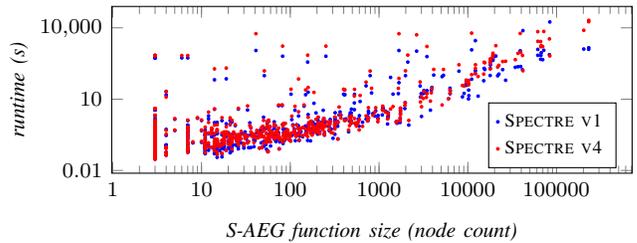

To show scalability of \tool{}, we use it to search for transient \textit{universal data transmitters} in the \libsodium{} crypto-library~\cite{libsodium}.
All experiments are run on an Intel(R) Xeon(R) Gold 6226R CPU @ 2.90GHz server featuring 4 processors, 16 cores per processor, and 512 GB of RAM.
Performance is summarized in Fig.~\ref{fig:libsodium-perf}.
Furthermore, we sacrifice completeness for performance by limiting \tool{}'s search space in two key ways. First, we allow at most one intermediate \texttt{rf} edge between subsequent \texttt{addr} dependencies (\S\ref{sec:leakage-engines}).
Second, we leverage a ``sliding window'' approach, in which for each candidate transmitter \tool{} only considers the set of instructions in the S-AEG that can reach the transmitter in $W_{size}$ instructions.
In practice, we believe these limitations do not significantly impact \tool{}'s ability to detect leakage. 
In particular, our sliding window approach excludes the discovery of universal data leakage in two key scenarios: 
(1) a live value sits in a register for a long time (greater than $W_{size}$ instructions), a condition that compiler register allocation tries to avoid; or (2) a {\it committed} instruction stores an unchecked value (e.g. an array pointer constructed using
an attacker-supplied index) to memory and does not read it for a long time.


We first analyze \libsodium{} using \tool{}'s \textsc{Spectre v1} engine to search for \S\ref{sec:modeling-leakage}'s universal data transmitter signature.
\tool{} assumes $d_\text{spec}=250$ and $W_{size}=500$.
Here, \tool{} reports many benign candidate executions featuring the pattern
\texttt{uint32\_t *idxp; \ldots = array[*idxp];}
where a
pointer is loaded from memory and subsequently dereferenced, forming the first \texttt{addr} dependency. While valid, this leakage is low-risk for two reasons. First, in bug-free code, a pointer (e.g. \texttt{idxp}) will rarely be
solely controlled by an attacker. Second, if the pointer is indeed not attacker-controlled, the same data will leak every time, so it acts as a \textit{non-universal} data transmitter.
To filter out these low-risk examples, we re-run \tool{}'s \textsc{Spectre v1} engine using a modified \texttt{addr} dependency chain (\texttt{addr\_gep.addr}).
Except for one false positive due to imprecise alias analysis, no universal data transmitters are found.
Recent work does not find any \textsc{Spectre v1} violations~\cite{ctfoundations:pldi:20} in \libsodium{}.
However, they restrict their search using taint annotations which likely confirms our hypothesis that the flagged data transmitters are benign.



Next, we
repeat the experiment above using
\tool{}'s \textsc{Spectre v4} engine, \textit{except} that
results are not filtered according to \texttt{addr\_gep}.
Since programs feature {\it many} more store forwarding speculation primitives (loads) than control-flow speculation primitives (branches), there is a larger search space with more complex constraints. Thus, we set \tool{}'s $d_\text{spec}=25$ with $W_{size} = 50$ to ensure analysis terminates. No universal data transmitters are found, corroborating recent work \cite{ctfoundations:pldi:20} which analyzes \libsodium{} with $d_\text{spec}=20$.



\tool{} employs various optimizations to scale to analyzing realistic-sized codebases. We omit a detailed description in our submission due to space constraints. Fig.~\ref{fig:libsodium-perf} shows that 93\%/92\% of 861 functions are analyzed in \textit{less then one minute} of serial execution time for \textsc{Spectre v1/v4}. \libsodium{} defines 943 functions, but \tool{} avoids re-analyzing functions which it has already inlined and checked elsewhere.

\section{Related Work}
\label{sec:related}

\textbf{Detecting Transient Leakage:}
Recent work simulates transient execution so that standard software analysis tools can detect classes of Spectre vulnerabilities in programs~\cite{oleksenko:specfuzz,guarnieri:spectector,ctfoundations:pldi:20,guanhua:kleespectre,guo:specusym,wu:abstractinterpretation}. \lcms{} take a similar approach via the \texttt{tfo} relation. 
SpecFuzz~\cite{oleksenko:specfuzz} and Spectector~\cite{guarnieri:spectector} detect \textsc{Spectre v1} gadgets in code using fuzzers and symbolic execution engines, respectively; Pitchfork~\cite{ctfoundations:pldi:20} uses symbolic execution to detect \textsc{Spectre v1/v1.1/v4} violations. \lcms{} also capture \textsc{Spectre v1.1} leakage, but we omit an example due to space limitations.

Researchers have proposed tools to detect Spectre-style vulnerabilities at the binary~\cite{2021:oo7,guarnieri:spectector,ctfoundations:pldi:20, 2021:huntingthehaunter} and LLVM-IR levels~\cite{guanhua:kleespectre,wu:abstractinterpretation,guo:specusym}. Nevertheless, all existing tools either scale poorly or face qualitative limitations.
For example, Spectector~\cite{guarnieri:spectector} only detects \textsc{Spectre v1} and does not scale well to large codebases. Spectector is also based on the program-counter security model~\cite{moldar:programcounter} and thus disallows branching on secrets.
\lcms{} support branching on secrets and are not limited to reasoning about vulnerabilities involving transient execution nor are they limited to capturing just a single Spectre variant.
Pitchfork~\cite{ctfoundations:pldi:20} detects \textsc{Spectre v1/v4\textsc{}}; however, its implementation is unsound~\cite{barthe:highassurance}, and its \textsc{Spectre v4} detection scheme scales poorly.

\textbf{Formalizing Transient Leakage:}
Recent research applies formal rigor to reasoning about the impact of transient execution attacks on software~\cite{barthe:highassurance,ctfoundations:pldi:20,patrignani2021exorcising,guarnieri:spectector,vassena:blade,guanciale:inspectre,guarnieri:contracts, oleksenko:specfuzz}. 
Cauligi et al.~\cite{ctfoundations:pldi:20} defines \textit{speculative constant-time} using an adversarial semantics for speculative execution. Similar to \lcms{}, their modeling approach captures a variety of transient execution attacks including \textsc{Spectre v1/v4}. 
InSpectre~\cite{guanciale:inspectre} features an operational model to support reasoning about transient execution attacks and countermeasures.
Guarnieri et al.~\cite{guarnieri:contracts} proposed hardware-software contracts to explicitly expose to software which aspects of microarchitectural state are observable to an adversary as a program executes.

Concurrent work~\cite{cat-spectre} also proposes to derive axiomatic security models from MCMs. Specifically, the authors use \texttt{.cat} models (\S\ref{sec:vocab}) out-of-the-box to formally model and automatically detect \textit{access instructions} in programs---namely memory read events which are capable of accessing secrets. In doing so, \texttt{rf} is used to model
\textit{both} architectural and microarchitectural data-flow.
This approach does not support modeling or identifying transmitters in programs and thus cannot determine if the data read by a particular access instruction can eventually be leaked via a transmitter.
Modeling micro-architectural leakage which does not involve architectural read events accessing secrets (as is the case for silent stores) is also not supported. 
In contrast, \lcms{} restrict \texttt{rf} to modeling architectural (software-visible) data-flow and introduce \texttt{rfx} to model microarchitectural data-flow. More generally, the architectural and microarchitectural semantics that 
\lcms{} define for programs enables the modeling and automatic detection of \textit{transmitters}. Whether a transmitter leaks \xstate{} or (universal) control/data via associated access instructions can be further deduced with the help of our transmitter taxonomy (\S\ref{sec:modeling-leakage}).


Just as \tool{} searches programs for transmitters, the work discussed above~\cite{cat-spectre} presents a tool \textsc{Kaibyo} which searches programs for access instructions capable of reading from a particular secret address. \textsc{Kaibyo} takes minutes (with a 90 minute timeout in some cases) to inspect the same benchmarks that \tool{} analyzes in less then as second.
Due to its focus on finding access instructions, \textsc{Kaibo} does not support optimal fence insertion.


Finally, Blade~\cite{vassena:blade} uses a static type system to eliminate transient leakage from CT cryptographic code. Blade prohibits speculative leakage by breaking flows from transiently-typed expressions to sinks with a hypothetical fence called \emph{protect}. 
Similar to Blade, \lcms{} can synthesize a minimum number of fences; they can also effectively use the \emph{protect} fence.
However, in contrast to Blade's conservative type system, \lcms{} are more accurate and lead to fewer false-positives. Further, while Blade's approach is limited to \textsc{Spectre v1}, \lcms{} capture different microarchitectural attacks and \tool{}'s current implementation supports fence insertion for both \textsc{Spectre v1/v4} \textit{without} transient types.
\section{Concluding remarks}
\label{sec:conclusions}
We propose \lcms{} as new security contracts that enable programmers, compiler writers, and runtime designers to reason about the security implications of hardware on software.
\lcms{} support precisely pinpointing hardware-related vulnerabilities in programs, as in \S\ref{sec:specexamples}. In turn, they support the design and development of (1) formal analysis frameworks, like \toolkit{} and (2) tools which can detect and repair vulnerable programs, like \tool{}.
Ultimately, we envision programmers using \lcms{} to specify security requirements by labeling data with \textit{trust domains}. A compiler can then translate a high-level programs to secure assembly code that prevents leakage across domains.

One limitation of \lcms{} is the type of side-channels they  capture---\lcms{} capture leakage that results from inter-instruction \textit{interactions} through hardware state rather than from operand-dependent variable time execution of individual instructions (e.g., due to subnormal floating point optimizations~\cite{andrysco:subnormal}). Such an enhancement to the formalism is left for future work.

\section*{Acknowledgements}
We would like to thank John Mitchell and Clark Barrett for their valuable discussions and feedback on this work.


\bibliographystyle{IEEEtranS}
\bibliography{refs}

\end{document}